\documentclass[preprint]{emulateapj}

\shortauthors{Fujii et al.}

\usepackage[breaklinks=true,colorlinks=true,linkcolor=blue,citecolor=blue,urlcolor=red,
setpagesize=false]{hyperref}
\usepackage{aas_macros}
\usepackage{amsmath}

\begin{document}
\title{Dense Molecular Clumps associated with the LMC Supergiant Shells LMC 4 \& LMC 5}

\author{Kosuke Fujii\altaffilmark{1,3}, Tetsuhiro Minamidani\altaffilmark{2}, Norikazu Mizuno\altaffilmark{1,3}, Toshikazu Onishi\altaffilmark{4}, Akiko Kawamura\altaffilmark{3}, Erik Muller\altaffilmark{3}, Joanne Dawson\altaffilmark{5,6}, 
Ken'ichi Tatematsu\altaffilmark{3}, Tetsuo Hasegawa\altaffilmark{3}, 
Tomoka Tosaki\altaffilmark{7}, Rie E. Miura\altaffilmark{3}, Kazuyuki Muraoka\altaffilmark{4},
Takeshi Sakai\altaffilmark{8}, Takashi Tsukagoshi\altaffilmark{9}, Kunihiko Tanaka\altaffilmark{10}, Hajime Ezawa\altaffilmark{3}, 
and Yasuo Fukui\altaffilmark{11}}

\altaffiltext{1}{Department of Astronomy, School of Science, The University of Tokyo, 7-3-1 Hongo, Bunkyo-ku, Tokyo 133-0033, Japan}
\altaffiltext{2}{Nobeyama Radio Observatory, 462-2 Nobeyama Minamimaki-mura, Minamisaku-gun, Nagano 384-1305}
\altaffiltext{3}{National Astronomical Observatory of Japan, 2-21-1 Osawa, Mitaka, Tokyo 181-8588, Japan}
\altaffiltext{4}{Department of Physical Science, Osaka Prefecture University, Gakuen 1-1, Sakai, Osaka 599-8531, Japan}
\altaffiltext{5}{Australia Telescope National Facility, CSIRO Astronomy and Space Science, PO Box 76, Epping, NSW 1710, Australia}
\altaffiltext{6}{Department of Physics and Astronomy and MQ Research Centre in Astronomy, Astrophysics and Astrophotonics, Macquarie University, NSW 2109, Australia}
\altaffiltext{7}{Joetsu University of Education, Yamayashiki-machi, Joetsu, Niigata 943-8512, Japan}
\altaffiltext{8}{Graduate School of Informatics and Engineering, The University of Electro-Communications, Chofu, Tokyo 182-8585, Japan}
\altaffiltext{9}{College of Science, Ibaraki University, Bunkyo 2-1-1, Mito 310-8512, Japan}
\altaffiltext{10}{Department of Physics, Faculty of Science and Technology, Keio University, 3-14-1 Hiyoshi, Yokohama, Kanagawa 223-8522, Japan}
\altaffiltext{11}{Department of Astrophysics, Nagoya University, Furo-cho, Chikusa-ku, Nagoya 464-8602, Japan}

\email{kosuke.fujii@nao.ac.jp}

\begin{abstract}
	We investigate the effects of Supergiant Shells (SGSs) and their interaction on dense molecular clumps by observing the Large Magellanic Cloud (LMC) star forming regions N48 and N49, which are located between two SGSs, LMC 4 and LMC 5. $^{12}$CO ($J$=3--2, 1--0) and $^{13}$CO($J$=1--0) observations with the ASTE and Mopra telescopes have been carried out towards these regions. A clumpy distribution of dense molecular clumps is revealed with 7 pc spatial resolution. Large velocity gradient analysis shows that the molecular hydrogen densities ($n({\rm H}_2)$) of the clumps are distributed from low to high density ($10^3$--$10^5$ cm$^{-3}$) and their kinetic temperatures ($T_{\rm kin}$) are typically high (greater than $50$ K). These clumps seem to be in the early stages of star formation, as also indicated from the distribution of H$\alpha$, young stellar object candidates, and IR emission. We found that the N48 region is located in the high column density \ion{H}{1} envelope at the interface of the two SGSs and the star formation is relatively evolved, whereas the N49 region is associated with LMC 5 alone and the star formation is quiet. The clumps in the N48 region typically show high $n({\rm H}_2)$ and $T_{\rm kin}$, which are as dense and warm as the clumps in LMC massive cluster-forming areas (30 Dor, N159). These results suggest that the large-scale structure of the SGSs, especially the interaction of two SGSs, works efficiently on the formation of dense molecular clumps and stars. 

\end{abstract}

\section{Introduction}
	Multiple stellar winds and supernova explosions from OB clusters form large-scale expanding structures known as superbubbles or supershells  \citep[for reviews, see][]{Tenorio-Tagle_etal_1988,Dawson_2013}. The interstellar medium (ISM) is shocked and swept up to form a dense, expanding shell, a process which strongly influences both the physical condition and the distribution of the ISM. The ISM of late-type galaxies often exhibits prominent, large shell structures with sizes approaching 1 kpc, which are called Supergiant shells (SGSs). 
In the Large Magellanic Cloud, SGSs were first identified as networks of ring-shaped of H$\alpha$ filaments, often with a large number of OB stars in their interior \citep{Meaburn_1980}. They are now more commonly identified as deficiencies in the neutral ISM bordered by regions of higher density, regardless of the presence of central OB clusters \citep{Kim_etal_1999,Walter_etal_1999,Book_etal_2008}. 
Several formation scenarios have been suggested for SGSs in nearby galaxies \citep{Dopita_etal_1985,Domgorgen_etal_1995,Braun_etal_1997,Efremov_etal_1998,Walter_etal_1999,Braun_etal_2000}, but all scenarios require input energy up to $\sim 10^{53}$ ergs, which is equivalent to the combined energy input of around 100 typical core collapse supernovae and the stellar winds of their progenitors. Such large shells may puncture the galactic gas disk and vent hot gas into the galactic halo \citep[e.g.,][]{MacLow_etal_1989}. 
Expanding shells are observed to become much larger in dwarf and irregular galaxies, where the differential rotation and galactic sheer are not significant. Therefore the shells in these galaxies grow to larger sizes (up to 1 kpc) and persist over longer timescales than in spiral galaxies \citep{McCray_Kafatos_1987,Walter_Brinks_1999}.

The compression of the interstellar medium (ISM) in Superbubbles is believed to trigger the formation of molecular clouds and star clusters \citep[e.g.,][]{McCray_Kafatos_1987,Elmegreen_1998,Ehlerova_etal_1997,Hartmann_etal_2001,Dawson_etal_2011b,Dawson_2013}. As the ambient diffuse ISM is swept up and collected, it forms a dense shell with high column density, which may become gravitationally unstable as it cools, eventually collapsing to form molecular clouds and stars (``collect and collapse''). Alternatively, the expansion of a shell could compress pre-existing dense molecular clouds around its edge, causing these clouds to collapse (``globule squeezing''). 
Previous observational studies have pointed out that SGSs, the largest structures formed by stellar feedback, do indeed trigger star formation at their rims \citep{Yamaguchi_etal_2001b,Book_etal_2009,Weisz_etal_2009}. 
It is also notable that the collision of SGSs is expected to drive violent episodes of star formation, and may be one process by which super-star clusters such as R136 in 30 Doradus are formed \citep{Chernin_etal_1995}. In addition, 2D models of colliding supershells suggest that small ($<$ 1 pc), dense (up to 10$^4$ cm$^{-3}$), cold ($<$ 100 K) gas clumps and filaments are formed naturally by thermal instabilities in the highly turbulent and compressed collisional zone \citep{Ntormousi_etal_2011}. Since global gravitational instabilities and/or spiral shocks are not effective drivers of star formation in low-mass galaxies, the impact of stellar feedback upon star formation and gas dynamics is considered to be more significant in such systems \citep[e.g.,][]{MacLow_Ferrara_1999}.

The Large Magellanic Cloud (LMC) is the nearest external galaxy to the Milky Way \citep[distance $\sim$ 50 kpc;][]{Pietrynski_etal_2013} and is relatively face-on to us \citep[inclination $\sim$ 35$^{\circ}$;][]{vanderMarel_Cioni_2001}. 
It also has a large population of SGSs and superbubbles in its gaseous disk \citep[e.g.,][]{Meaburn_1980,Chu_MacLow_1990,Kim_etal_1999}. Previous work has suggested that molecular cloud formation is enhanced around LMC SGSs \citep{Yamaguchi_etal_2001b, Dawson_etal_2013}, and recent star formation occurs preferentially along their  peripheries \citep{Yamaguchi_etal_2001b,Book_etal_2009}. These studies have discussed molecular cloud and star formation on the size scales of giant molecular clouds \citep[GMCs; size $\sim$ 10--100 pc, mass $\sim$ 10$^5$--10$^7$ M$_{\odot}$;][]{Fukui_etal_2008}. 
Observational studies of molecular ``clumps'' (density $>10^3$ cm$^{-3}$, size $\sim$1--10 pc, mass $\sim$10$^4$ M$_{\odot}$) in the LMC \citep{Minamidani_etal_2008, Minamidani_etal_2011} have shown that their densities and temperatures increase generally from those in GMCs with no signs of massive star formation, to those in GMCs with young clusters. This difference seems to reflect an evolutionary trend in the star formation activity in molecular clumps, and observational studies probing the size scales of molecular clumps are therefore a useful means of understanding the evolution of GMCs. It is important to investigate the characteristics of dense molecular clumps under the influence of SGSs, in order to study the relation between their physical properties and the global structure of the shells.

\begin{figure}
\epsscale{.80}
\plotone{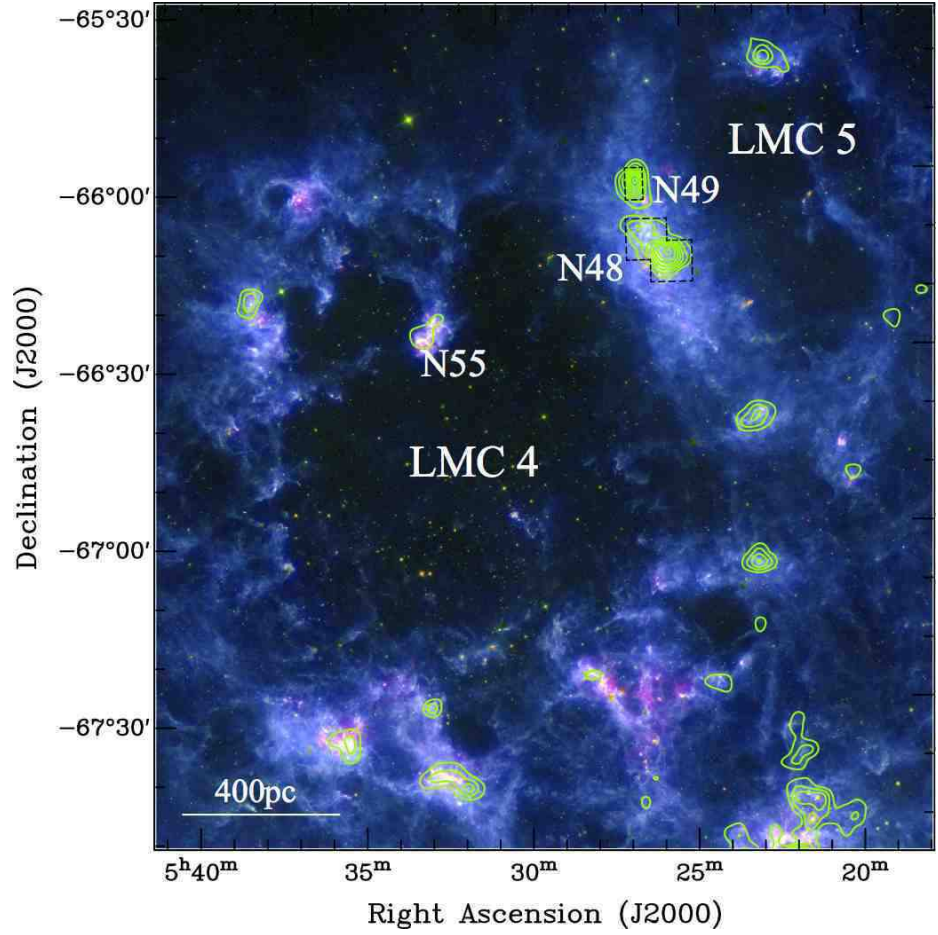}
\caption{Spitzer three-color image \citep[R:24$\mu$m, G:8.0$\mu$m, B:160$\mu$m;][]{Meixner_etal_2006}of the LMC 4/LMC 5 region. Green contours are NANTEN $^{12}$CO($J$=1--0) integrated intensity \citep{Fukui_etal_2008}. The contours start at 1.2 K km s$^{-1}$ and are incremented in steps of 1.2 K km s$^{-1}$. Black dashed lines indicate the region observed with ASTE. \label{fig_LMC45_CO}}
\end{figure}

In this study, we focus on the star forming regions N48 and N49 in the LMC \citep{Henize_1956}, which are located between two SGSs, LMC 4 and LMC 5 \citep[][see also Figure \ref{fig_LMC45_CO}]{Meaburn_1980}. This situation is similar to 30 Doradus, including the R 136 cluster, which is located toward the interface between two SGSs, LMC 2 and LMC 3 \citep{Tenorio-Tagle_etal_1988}. Both LMC 4 and LMC 5 consist of diffuse H$\alpha$ filaments and bright \ion{H}{2} regions \citep{Meaburn_1980}, and giant holes in the \ion{H}{1} gas \citep{Dopita_etal_1985,Kim_etal_1999} and interstellar dust \citep{Meixner_etal_2006}. LMC 4 is the largest SGS in the LMC, with a size of $1.0 \times 1.8$ kpc in H$\alpha$ \citep{Meaburn_1980}, and LMC 5 is located northwest of LMC 4 with a diameter of $\sim$ 800 pc in H$\alpha$ \citep{Meaburn_1980}. Of all SGSs in the LMC, the LMC4/LMC5 region shows some of the strongest evidence for the enhanced formation of molecular clouds due to the action of the shells on the ISM \citep{Dawson_etal_2013}. In the N48, N49 regions, two massive GMCs whose total mass is $\sim$1.5$\times 10^{6}$ M$_{\odot}$ have been identified with the NANTEN telescope (catalogued alternatively as LMC/M5263--6606 and LMC/M5253--6618 \citep{Mizuno_etal_2001}, or LMC N J0525--6609 \citep{Fukui_etal_2008}). Since they correspond to 50\% of the total mass of molecular clouds associated with LMC 4 \citep{Yamaguchi_etal_2001a},  and the region shows a high ratio of molecular to atomic gas 
\citep[$\sim$60\%;][]{Mizuno_etal_2001}, they are considered to have formed efficiently within the dense \ion{H}{1} ridge swept up by the two SGSs. However, they contain small \ion{H}{2} regions (N48) and a single SNR (N49), both of which are located at the peripheries of the GMCs, therefore they are considered to be early-stage cluster-forming clouds \citep{Mizuno_etal_2001}. In addition, \cite{Cohen_etal_2003} have identified a large, dense ridge-shaped photodissociation region that lies between the two SGSs, and argued that the feature is a strong candidate for secondary star formation by the interaction of two SGSs. These regions are therefore an excellent target in which to investigate how the action of SGSs and their interaction affect the physical properties of dense molecular clumps.

New $^{12}$CO($J$=3--2, 1--0) and $^{13}$CO($J$=1--0) line mapping observations towards the N48 and N49 regions were made for this purpose. The details of the observations are described in \S 2. In \S 3, the $^{12}$CO($J$=3--2) intensity distribution and physical properties of the clumps are presented. In \S 4, comparisons of the CO distribution with star formation tracers such as \ion{H}{2} regions and young stellar objects are shown, and relations between the physical properties of the molecular clumps and the star formation activity in and around them are discussed. In \S 5, we compare the CO distribution with \ion{H}{1} gas to investigate the relation between the global structure of the SGSs and the physical parameters of the clumps. We also discuss the timescales of triggering events and finally summarize our scenario for evolutionary processes at work in the N48/N49 region.

\section{Observations}
	\subsection{$^{12}$CO($J$=3--2) observations}
Observations of the $^{12}$CO($J$=3--2) transition towards the N48 and N49 regions were made with the ASTE 10 m telescope at Pampa la Bola in Chile \citep{Ezawa_etal_2004}, in September 2006 and September 2011, respectively. The half-power beam width was measured to be 22$^{\prime \prime}$ at 345 GHz by observing the planets. Both observations were performed in the On-the-fly (OTF) mapping mode.

In 2006, we observed the giant molecular cloud LMC/M5263--6606 \citep{Mizuno_etal_2001} in the N49 region. The size of the $^{12}$CO($J$=3--2) mapping area was 3.0$^{\prime}$ $\times$ 5.5$^{\prime}$ (45 $\times$ 83 pc), which included the whole cloud. Along each row of an OTF field, individual spectra were recorded every 1.5$^{\prime \prime}$ and the spacing between the rows was 6$^{\prime \prime}$, so that the 22$^{\prime \prime}$ beam of the telescope was oversampled. The OTF mapping was performed along two orthogonal directions, i.e., scans along the right ascension and declination directions. In this period, we used a single cartridge-type double-side-band (DSB) SIS receiver, SC345 \citep{Kohno_2005}. The spectrometer was an XF-type digital autocorrelator MAC \citep{Sorai_etal_2000} and was used in the wide-band mode, which has a bandwidth of 512 MHz with 1024 channels. The spectrometer provided velocity coverage and channel spacing of 450 and 0.44 km s$^{-1}$ at 345 GHz, respectively. The chopper-wheel technique was employed to calibrate the antenna temperature $T_{\rm a}^{\ast}$. The typical system noise temperature during the observation was $\sim$500 K in DSB including atmospheric effects. The pointing error was measured to be within 5$^{\prime \prime}$ (peak to peak) by observing the CO point sources R Dor ($\alpha _{\rm B1950}$ = 4h 36m 45.84s, $\delta _{\rm B1950}$ = $-$62$^{\circ}$ 04$^{\prime}$ 35.70$^{\prime \prime}$) or o Cet ($\alpha _{\rm B1950}$ = 2h 19m 20.8s, $\delta _{\rm B1950}$ = $-$2$^{\circ}$ 58$^{\prime}$ 40.70$^{\prime \prime}$) every two hours during the observing period. We observed N159W ($\alpha _{\rm B1950}$ = 5h 40m 3.7s, $\delta _{\rm B1950}$ = $-$69$^{\circ}$ 47$^{\prime}$ 00.0$^{\prime \prime}$) every two hours to check the stability of the intensity calibration. The average and the standard deviation of the antenna temperature $T_{\rm a}^{\ast}$ of N159W was 5.48 $\pm$ 1.42 K, i.e. the intensity variation during these observations was estimated to be less than 26\%. We scaled the observational data taken in 2006 to $T_{\rm mb}$ scale with a scaling factor of 2.538 so that the average value of observed $T_{\rm a}^{\ast}$ of N159W is consistent with the main beam temperature of N159W $T_{\rm mb}$ = 13.9 $\pm$ 0.7 K, which was estimated by \cite{Minamidani_etal_2011}.

In 2011, we observed the giant molecular cloud LMC/M5253--6618 \citep{Mizuno_etal_2001} in the N48 region. 
The cloud was covered by two 7$^{\prime}$ $\times$ 7$^{\prime}$ (105 $\times$ 105 pc) square OTF maps. Along each row of an OTF field, individual spectra were recorded every 1.6$^{\prime \prime}$ and the spacing between the rows is 7.5$^{\prime \prime}$, so that the 22$^{\prime \prime}$ beam of the telescope is oversampled. The OTF mapping was performed both along the right ascension and declination directions, respectively. In this period, we used the waveguide-type sideband-separating SIS mixer receiver for single side band (SSB) operation, CATS345 \citep{Ezawa_etal_2008,Inoue_etal_2008}. The image rejection ratio at 345 GHz was estimated to be $\sim$ 10 dB. The spectrometer was an XF-type digital autocorrelator MAC \citep{Sorai_etal_2000} and was used in the high-resolution mode, which has a bandwidth of 128 MHz with 1024 channels. 
The velocity coverage and the channel spacing at 345 GHz were 125 and 0.11 km s$^{-1}$, respectively. Typical system noise temperatures during the observation were 500 K in SSB including atmospheric effects. The pointing error was measured to be within 5$^{\prime \prime}$ by observing a CO point source R Dor every two hours. We observed N159W every two hours to check stability, and the average and standard deviation of the antenna temperature was found to be 7.58 $\pm$ 0.26 K on a $T_{\rm a}^{\ast}$ scale. The intensity variation was therefore estimated to be less than 3\%. We scaled the observational data in 2011 to $T_{\rm mb}$ scale with scaling factor of 1.835.

The data reduction was made using the software package NOSTAR, which comprises tools for OTF data analysis, developed by the National Astronomical Observatory of Japan \citep{Sawada_etal_2008}. Linear baselines were subtracted from the spectra. The raw data were re-gridded to 10$^{\prime \prime}$ per pixel, giving an effective spatial resolution of approximately 27$^{\prime \prime}$, which corresponds to 7 pc at the distance of the LMC. The data sets taken along the right ascension and declination directions were co-added by the Basket-weave method \citep{Emerson_Graeve_1988} to remove any effects of scanning noise. A fifth order polynomial function was then fitted to the baseline and subtracted in order to reduce the effects of baseline ripples. The 2011 data was binned to a channel spacing of 0.44 km s$^{-1}$, which corresponds to the channel spacing of the MAC wide-band mode used in the 2006 data.

\subsection{$^{12}$CO($J$=1--0) and $^{13}$CO($J$=1--0) observations}
We performed observations of the $^{12}$CO($J$=1--0) and $^{13}$CO($J$=1--0) transitions using the 22 m Australia Telescope National Facility (ATNF) Mopra telescope in June/July 2012. We observed eleven 2$^{\prime}$ $\times$ 2$^{\prime}$ areas which covered the prominent parts of the $^{12}$CO($J$=3--2) clumps. The half-power beam width was 33$^{\prime \prime}$ at the 3 mm band, and the observations were performed in the OTF mapping mode. Individual spectra were recorded every 6$^{\prime \prime}$ and the spacing between rows is 9$^{\prime \prime}$, so that the 33$^{\prime \prime}$ (FWHM) telescope beam was oversampled.

We used the 3mm MMIC receiver that can simultaneously record dual polarization data. The spectrometer was the Mopra Spectrometer (MOPS) digital filter bank and was used in the zoom-band mode, which can record up to sixteen 137.5 MHz zoom bands positioned within an 8 GHz window. The spectrometer provided a velocity coverage and channel spacing of 376 km s$^{-1}$ and 0.09 km s$^{-1}$ at 3 mm, respectively. 
Typical system noise temperatures during the observations were 600 K at the frequency of the$^{12}$CO line, and 250 K at the frequency of the $^{13}$CO line, including atmospheric effects. The pointing error was measured to be within 10$^{\prime \prime}$  by observing the SiO maser toward R Dor ($\alpha _{\rm J2000}$ = 4h 36m 45.61s, $\delta _{\rm J2000}$ = $-$62$^{\circ}$ 04$^{\prime}$ 37.92$^{\prime \prime}$) every one to two hours during the observing period. We observed Orion KL ($\alpha _{\rm J2000}$ = 5h 35m 14.5s, $\delta _{\rm J2000}$ = $-$5$^{\circ}$ 22$^{\prime}$ 29.56$^{\prime \prime}$) once a day to check the stability of the intensity calibration. The average and the standard deviation of Orion KL antenna temperatures was 47.9 $\pm$ 3.1 K in $T_{\rm a}^{\ast}$ scale, then the intensity variation during the observation was estimated to be less than 7\%. The efficiencies we have assumed for CO are the ``extended beam efficiency'' $\eta _{\rm xb}$ discussed by \cite{Ladd_etal_2005}, which includes the effect of coupling to the inner error beam for sources larger than $\sim$ 2$^{\prime}$. We determined $\eta _{\rm xb} = 0.48$ by dividing the observed the peak antenna temperature by 100 K, which is the corrected peak CO main beam temperature for Orion KL given by \cite{Ladd_etal_2005}. 
In this paper, data reduction was performed using the ATNF's ${\it Livedata}$, and ${\it Gridzilla}$ software packages. ${\it Livedata}$ performs bandpass calibration using off-source spectra, then fits and subtracts a linear baseline. ${\it Gridzilla}$ takes the spectra and grids them onto a data cube using a Gaussian smoothing kernel of FWHM 33$^{\prime \prime}$, comparable to that of the Mopra primary beam. We weighted the spectra by the inverse of the system temperature when gridding. The resulting effective spatial resolution is approximately 45$^{\prime \prime}$, which corresponds to 11 pc at the distance of the LMC. Then we fit and subtracted a fifth order polynomial function from the baseline to reduce effects of baseline ripples. The channel width was binned up to 0.44 km s$^{-1}$, which matches the channel spacing of ASTE data. Finally, we re-gridded to a right ascension and declination grid with a spacing of 10$^{\prime \prime}$ (to match as the ASTE data) using the ATNF's ${\it MIRIAD}$ software.
All parameters of the ASTE and Mopra observations are summarized in Table \ref{tab_obs}.

\section{Results}
	We first present the spatial distributions of the $^{12}$CO($J$=3--2) emission from the molecular clouds at 7 pc resolution, and the $^{12}$CO($J$=1--0) and $^{13}$CO($J$=1--0) emission at 11 pc resolution (\S 3.1). We then identify molecular clumps and estimate the physical parameters of each clump (\S 3.2, \S 3.3).

\subsection{Spatial distributions of CO transitions}
\begin{figure}
\epsscale{1.0}
\plotone{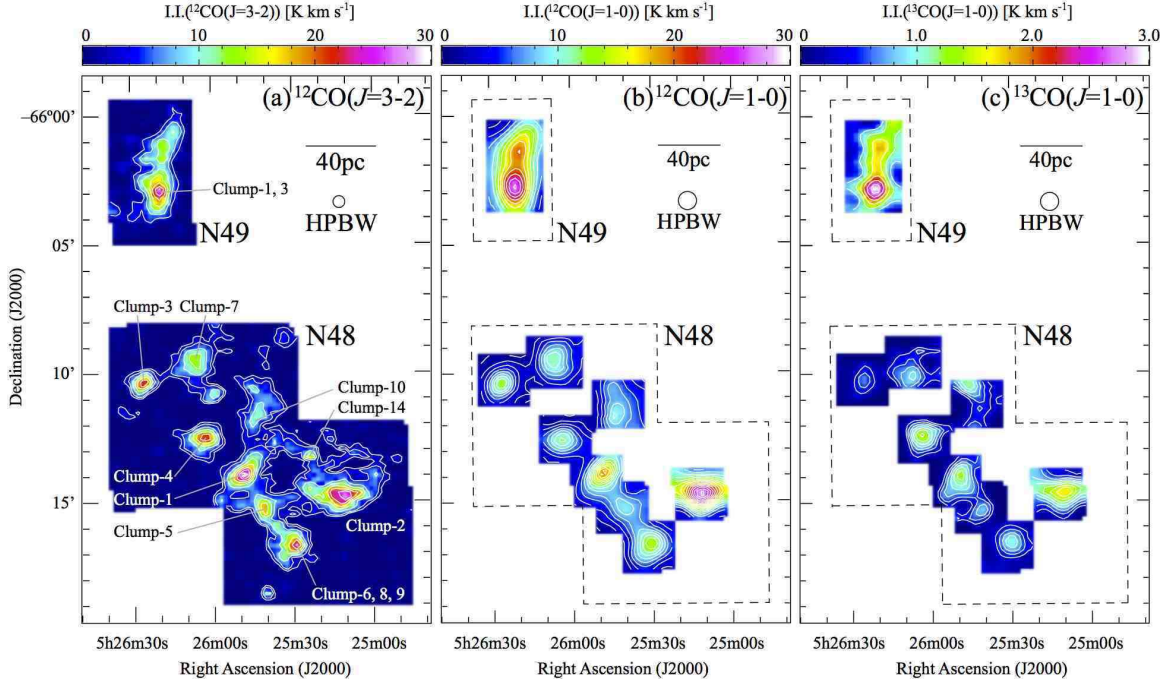}
\caption{(a) $^{12}$CO($J$=3--2) integrated intensity map of the N49 (north) and N48 (south) regions. In the N49 region, the integration range is 279.8 to 300.1 km s$^{-1}$, and the contour levels are 3.4, 6.8, 10.8, 14.8, 18.8, 22.8, 26.8 K km s$^{-1}$. The first two contour levels correspond to 5$\sigma$ and 10$\sigma$ noise levels, and thereafter run in steps of 4.0 K km s$^{-1}$. In the N48 region, the integration range is 275.0 to 310.1 km s$^{-1}$, and the contour levels are 2.0 (5$\sigma$), 4.0, 8.0, 12.0, 16.0, 20.0, 24.0, and 28.0 K km s$^{-1}$. The circle shows the ASTE effective beam size (27$^{\prime \prime}$). Several Clump ID numbers defined in \S\ref{Clump_def} are also labelled on the figure. (b) $^{12}$CO($J$=1--0) integrated intensity maps obtained with Mopra. Black dashed lines indicate the area observed with ASTE. The integration ranges are the same as in (a). In the N49 region, the lowest contour is 1.2 K km s$^{-1}$ (the 5 $\sigma$ noise level), and thereafter run in steps of 2.4 K km s$^{-1}$. In the N48 region, the lowest contour is 0.75 K km s$^{-1}$ (5 $\sigma$) and thereafter run in steps of 1.5 K km s$^{-1}$. The circle is the Mopra effective beam size (45$^{\prime \prime}$). (c) $^{13}$CO($J$=1--0) integrated intensity maps. The integration ranges are the same as for (a). In the N49 region, the lowest contour is 0.78 K km s$^{-1}$ (5 $\sigma$), and thereafter run in steps of 0.39 K km s$^{-1}$. In the N48 region, the lowest contour is 0.34 K km s$^{-1}$ (5 $\sigma$) and thereafter run in steps of 0.17 K km s$^{-1}$. The circle is the Mopra effective beam size. \label{fig_COmaps} }
\end{figure}

Figure \ref{fig_COmaps}(a) shows a $^{12}$CO($J$=3--2) integrated intensity map of the N48 and N49 regions obtained with ASTE. Typical rms noise levels are 0.68 K km s$^{-1}$ and 0.40 K km s$^{-1}$ in the N49 and N48 region respectively. 
\cite{Minamidani_etal_2008} have also performed similar resolution and sensitivity observations with ASTE toward GMCs in the LMC, which are located in the molecular ridge extending southward from 30 Doradus, and the ``CO Arc'' along the southeastern optical edge. Compared to these GMCs, the distribution of $^{12}$CO($J$=3--2) clouds in the N48 and N49 regions tends to be more clumpy; that is each dense molecular clump tends to be independently distributed and not be surrounded by a diffuse gas envelope.  

Figure \ref{fig_COmaps}(b) shows a $^{12}$CO($J$=1--0) integrated intensity map of the N48 and N49 regions obtained with Mopra. Typical rms noise levels are 0.24 K km s$^{-1}$ in the N49 region, and 0.15 K km s$^{-1}$ ($^{12}$CO) in the N48 region. 
Figure \ref{fig_COmaps}(c) shows the equivalent $^{13}$CO($J$=1--0) map. Typical rms noise levels are 0.16 K km s$^{-1}$ in the N49 region, and 0.068 K km s$^{-1}$ in the N48 region. Although the spatial resolution is lower than the ASTE observations, the $^{13}$CO($J$=1--0) distribution traces well the bright features seen in $^{12}$CO($J$=3--2), whereas the $^{12}$CO($J$=1--0) distribution is more extended.

\begin{figure}
\epsscale{.80}
\plotone{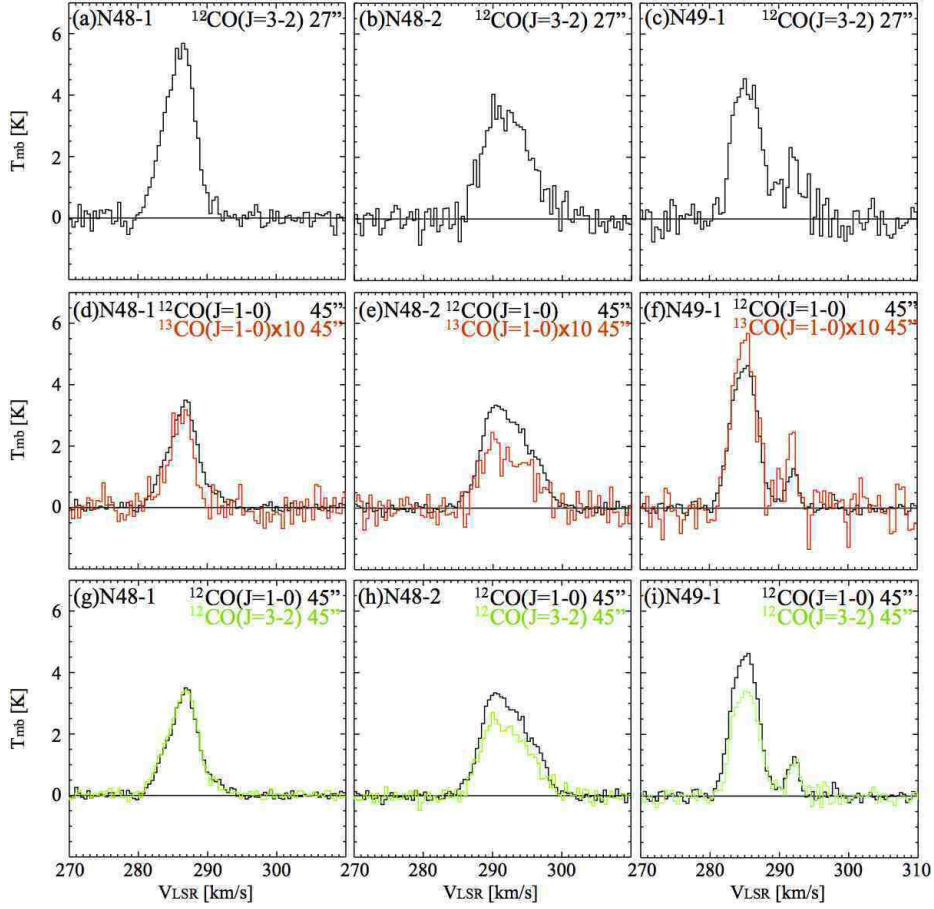}
\caption{Typical peak line profiles of the observed molecular clouds (N48-1, N48-2, N49-1, see also Table \ref{tab_peak}). (a)(b)(c): $^{12}$CO($J$=3--2) profiles obtained with ASTE. (d)(e)(f): $^{12}$CO($J$=1--0) (black) and $^{13}$CO($J$=1--0) (red) profiles obtained with Mopra.  (g)(h)(i): $^{12}$CO($J$=1--0) profiles (black) and $^{12}$CO($J$=3--2) profiles smoothed to the same (45$^{\prime \prime}$) resolution (green). Note that the vertical scales are main-beam temperature, $T_{\rm mb}$, but $^{13}$CO profiles are multiplied by a factor of ten. The velocity channel widths of all profiles are 0.44 km s$^{-1}$. \label{fig_profile}}
\end{figure}

In Figure \ref{fig_profile}, typical peak $^{12}$CO($J$=3--2), $^{12}$CO($J$=1--0) and $^{13}$CO($J$=1--0) line profiles are presented. The positions selected are local $^{12}$CO($J$=3--2) integrated intensity peaks (See \S \ref{Clump_def} and Table \ref{tab_peak}). The upper panels, Figures \ref{fig_profile}a--\ref{fig_profile}c, show the $^{12}$CO($J$=3--2) profiles. These illustrate the typical noise level of the data, $\sim$0.4 K rms at 0.44 km s$^{-1}$ channel width. The $^{12}$CO($J$=3--2) intensity is brightest at N48-1 ($T_{\rm mb}$ $\sim$5.5 K). The line profiles are in general single peaked, but some of them show two clearly-separated components, as illustrated in \ref{fig_profile}c. Figures \ref{fig_profile}d-\ref{fig_profile}f show the $^{12}$CO($J$=1--0) and $^{13}$CO($J$=1--0) profiles toward the same positions. The peak velocities and line widths of $^{13}$CO($J$=1--0) are similar to those of $^{12}$CO($J$=1--0).
Figures \ref{fig_profile}g-\ref{fig_profile}i show the $^{12}$CO($J$=1--0) and $^{12}$CO($J$=3--2) spectra together, in which the latter have been extracted from data smoothed to the same effective resolution as the Mopra data ($\sim45"$). The shapes of the $^{12}$CO($J$=3--2) profiles are similar to $^{12}$CO($J$=1--0), and nearly the same as $^{13}$CO($J$=1--0). This indicates that these lines trace similar parts of the clouds.

\subsection{$^{12}$CO($J$=3--2) Velocity Structure}
\begin{figure}
\epsscale{1.0}
\plotone{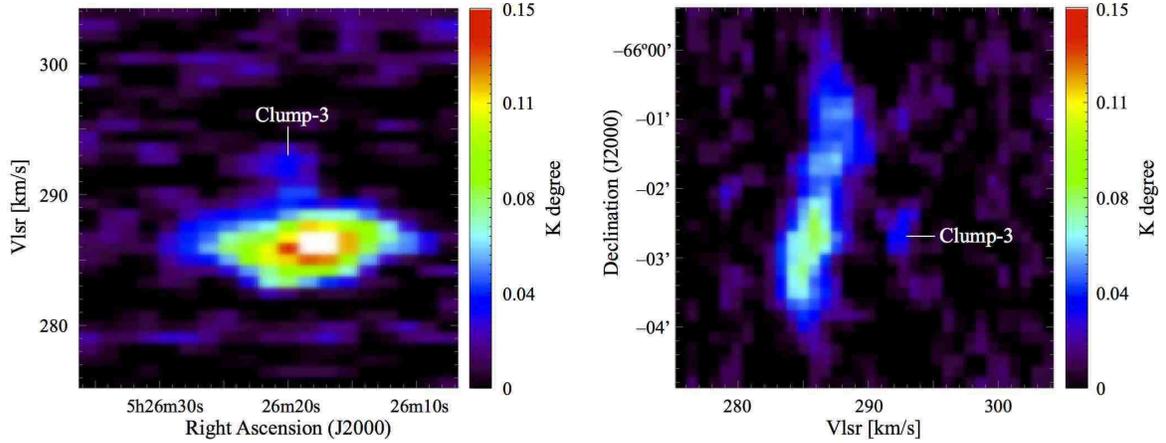}
\caption{Right Ascension-velocity diagram (left) and Declination-velocity diagram (right) of the $^{12}$CO($J$=3--2) emission in the N49 region. \label{fig_PVN49}}
\end{figure}

\begin{figure}
\epsscale{1.0}
\plotone{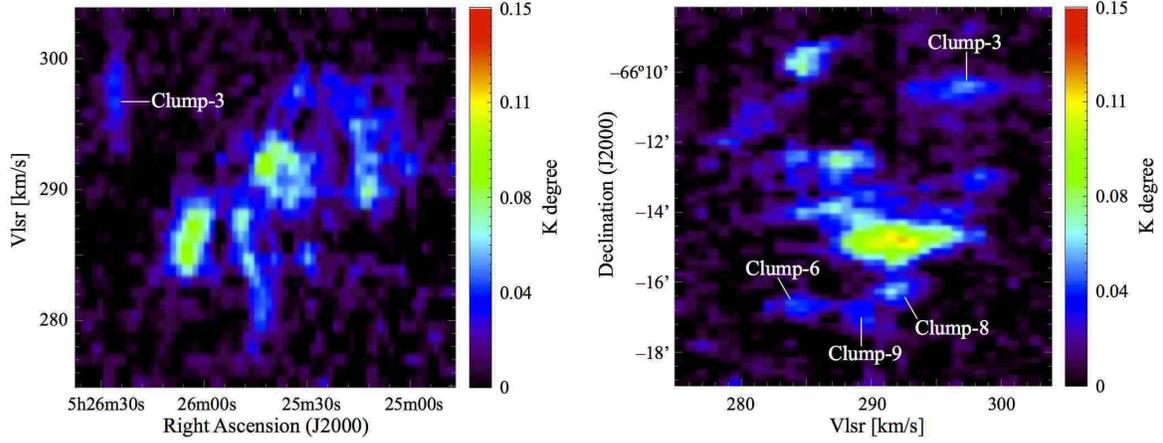}
\caption{The same position-velocity diagrams for the N48 region. \label{fig_PVN48}}
\end{figure}

Figures \ref{fig_PVN49} and \ref{fig_PVN48} show position-velocity diagrams of the $^{12}$CO($J$=3--2) emmission in both right ascension and declination, i.e. the cube collapsed along each spatial axis. In these diagrams, the velocity pixel width were bound up to 0.88 km s$^{-1}$. 

In the N49 region, $^{12}$CO($J$=3--2) emission is distributed between $\sim$283--290 km s$^{-1}$. The overall velocity structure of the cloud shows moderate velocity gradients. There is a compact second component around $(\alpha _{\rm J2000}, \delta _{\rm J2000}) \sim$ (5h 26m 20s, $-66^{\circ}$ 2$^{\prime}$ 40$^{\prime \prime}$), and $V_{\rm LSR} \sim 292$ km s$^{-1}$, which is defined as clump N49--3 (see \S \ref{Clump_def}). In the N48 region, $^{12}$CO($J$=3--2) emission is distributed between $\sim$277--300 km s$^{-1}$, and shows relatively clumpy structure. The overall velocity structure of the cloud has velocity gradients in both right ascension and declination. 
There is an isolated component around $(\alpha _{\rm J2000}, \delta _{\rm J2000}) \sim$ (5h 26m 26s, $-$66$^{\circ}$ 10$^{\prime}$ 24$^{\prime \prime}$), and $V_{\rm lsr} \sim 298$ km s$^{-1}$ (N48--3). The south clump located around $\delta _{\rm J2000} \sim -$66$^{\circ}$ 16$^{\prime}$ 30$^{\prime \prime}$, is broken into several velocity components (N48--6, 8, and 9).

\subsection{Properties of the $^{12}$CO($J$=3--2) clumps \label{Clump_def}}
To discuss the physical properties of dense molecular clumps, we define $^{12}$CO($J$=3--2) clumps in the way described in \cite{Minamidani_etal_2008}. The clump identification criteria are as follows: (1) Identify local peaks in the integrated intensity map that are greater than the 10 $\sigma$ noise level ($\geq$4.0 K km s$^{-1}$ for the N48 region, and $\geq$6.8 K km s$^{-1}$ for the N49 region). (2) Select those local peaks that have a peak brightness temperatures greater than the 3 $\sigma$ noise level of the  spectrum (typically $\sim$0.90 K for the N48 region, and  $\sim$1.0 K for the N49 region), then draw a contour at one-half of the peak integrated intensity level and identify it as a clump unless it contains other local peaks. (3) When there are other local peaks inside the contour, draw new contours at the 70\% level of each integrated intensity peak. Then, identify clumps separately if their contours do not contain another local peaks (the boundary is taken at the minimum integrated intensity between the peaks), or else identify a clump using the contour of the highest peak as a clump boundary. (4) if a spectrum has multiple velocity components with a separation of more than 5 km s$^{-1}$, identify those components as being associated with different clumps. 

Using these criteria, we identify 18 clumps in the N48 region and 3 clumps in the N49 region. The observed line parameters at the local peaks are shown in Table \ref{tab_peak}. The parameters of the identified clumps are listed in Table \ref{tab_clump}. Deconvolved clump sizes, $R_{\rm deconv}$, are defined as $[R^2 _{\rm nodeconv} - (\theta _{\rm HPBW}/2) ^2 ]^{0.5}$. $R_{\rm nodeconv}$ is the effective radius defined as $(A/\pi)^{0.5}$, where $A$ is the observed total cloud surface area. $V _{\rm LSR}$ and the composite line width, $\Delta V _{\rm clump}$, are derived using a single Gaussian fit to the spectrum obtained by averaging all spectra within a single clump. 

The virial mass is estimated as
\begin{equation}
\label{eq:m_vir}
M_{\rm vir} [{\rm M}_{\odot}] = 190\; \Delta V_{\rm clump}^{2}[{\rm km\;s^{-1}}]\; R[{\rm pc}], 
\end{equation}
which assumes that the clumps are spherical with density profiles of $\rho \propto r^{-1}$, where $\rho$ is the number density and $r$ is the distance from the cloud center \citep{MacLaren_etal_1988}. 

The $^{12}$CO($J$=3--2) luminosity of the cloud $L_{{\rm CO}(J=3-2)}$ is the integrated flux scaled by the square of the distance, 
\begin{equation}
L_{\rm CO}\; [{\rm K\;km\;s}^{-1}\;{\rm pc}^2] = D^2 \left( \sum T_{i} \right) \delta v \delta x \delta y, 
\end{equation}
where $T_ {i}$ is the $^{12}$CO($J$=3--2) brightness temperature of an individual voxel, $D$ is the distance to the LMC in parsecs (taken to be 5 $\times$ 10$^4$), $\delta x$ and $\delta y$ are the angular pixel dimensions in radians, and $\delta v$ is the width of one channel in km s$^{-1}$.

\begin{figure}
\epsscale{1.0}
\plotone{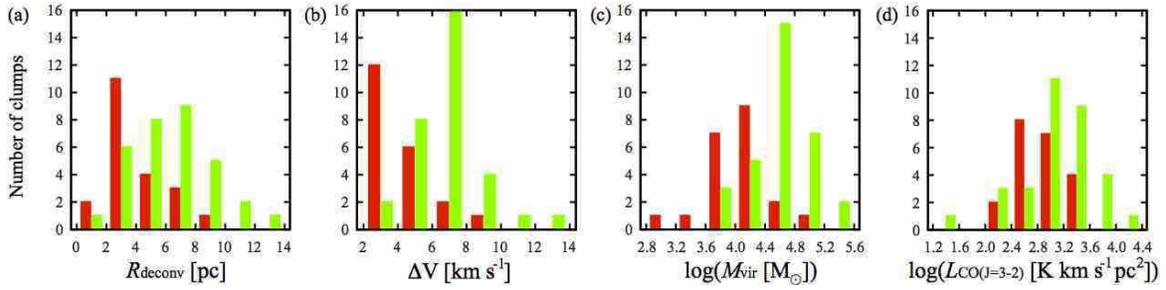}
\caption{Histograms of the physical properties of the $^{12}$CO($J$=3--2) clumps: (a) deconvolved sizes, (b) FWHM line widths, (c) virial masses, and (d) $^{12}$CO($J$=3--2) luminosities of the clumps (see text for definitions). Red and green bars correspond to the clumps in the N48/N49 region and the clumps of \citep{Minamidani_etal_2008}, respectively. \label{fig_hist}}
\end{figure}

Histograms of the $^{12}$CO($J$=3--2) clump physical properties are presented in Figure \ref{fig_hist}.  
Mean values in the N48 and N49 regions are clump sizes of $R_{\rm deconv} \sim 4.7$ pc, line widths of $\Delta V_{\rm clump} \sim 4.3$ km s$^{-1}$, virial masses of $M_{\rm vir} \sim 2.0 \times 10^{4}$ M$_{\odot}$, and $^{12}$CO($J$=3--2) luminosities of $L_{{\rm CO}(J=3-2)}\sim 1.2 \times 10^3$ K km s$^{-1}$ pc$^2$. Note that the smallest clump identified by the procedure above is detected in seven spatial pixels, and its deconvolved radius is 1.2 pc. For comparison, the physical properties of $^{12}$CO($J$=3--2) clumps identified in \cite{Minamidani_etal_2008} are also shown. 
These clumps are identified in GMCs located in the more active star forming regions of the LMC (30 Dor, N159, N166, N171, N206, N206D, GMC225; in the molecular ridge and the CO Arc), and the mass of these GMCs ranges between 0.4  to 10 $\times 10^6$ M$_{\odot}$ \citep[30 Dor $\sim 0.7 \times 10^6$ M$_{\odot}$, N159 and N171 $\sim 10 \times 10^6$ M$_{\odot}$, N166 $\sim 2 \times 10^6$ M$_{\odot}$, N206 $\sim 0.9 \times 10^6$ M$_{\odot}$, N206D $\sim 0.4 \times 10^6$ M$_{\odot}$, GMC 225 $\sim 1 \times 10^6$ M$_{\odot}$;][]{Fukui_etal_2008}. Although the mass of the GMC in the N48/N49 region ($\sim 1.5 \times 10^6$ M$_{\odot}$) is comparable to these GMCs, the physical properties of the clumps in the N48/N49 region (size, linewidth, virial mass and luminosity) all tend to be smaller, and the number of identified clumps in the N48 region is notably large (18 for N48, 3 for N49, 5 for 30 Dor, 16 for N159 and N171, 5 for N166, 2 for N206, 1 for N206D, and 3 for GMC 225). This suggests that the GMC in the N48/N49 region consists of a concentration of compact clumps -- a fact which was also indicated by the very clumpy distribution of the $^{12}$CO($J$=3--2) emission.

\subsection{Clump properties derived from the $^{13}$CO($J$=1--0) transition \label{sec_13COclump}}
Since $^{13}$CO is assumed to be optically thin for a typical molecular cloud, column density can be obtained from $^{13}$CO luminosity under the assumption of local thermodynamic equilibrium. This gives a better estimate of clump mass than with $^{12}$CO, which is typically optically thick. 
In order to compare luminosity-based mass with the virial mass, we also derive the physical properties of the clumps using the $^{13}$CO($J$=1--0) transition. However, the signal to noise ratio of $^{13}$CO($J$=1--0) is not sufficient to identify all clumps via step 1 of the clump identification method described above, i.e. the peak integrated intensity of several clumps is weaker than the 10 $\sigma$ noise level. We instead use the $^{12}$CO($J$=3--2) clump boundaries to derive the $^{13}$CO($J$=1--0) clump properties, supposing that the $^{13}$CO($J$=1--0) distribution is similar to that of $^{12}$CO($J$=3--2). Since the spatial distributions and the line profiles of the two transitions are similar (Fig. \ref{fig_COmaps}(a), (b), \ref{fig_profile}), this is likely to be a reasonable assumption. We identify those clumps with $^{13}$CO($J$=1--0) spectral peaks greater than the 3 $\sigma$ noise level at the $^{12}$CO($J$=3--2) local peak positions. The selected clumps are N48--1, 2, 3, 4, 5, 6, 7, 8, 10, 14, N49--1, 2, 3, and their physical properties are listed in table \ref{tab_13CO}. $T_{\rm mb}$, $V_{\rm LSR}$, $\Delta V$ were derived in the same way as for the $^{12}$CO($J$=3--2) clumps. Virial masses $M_{\rm vir}$ were estimated from eq.(\ref{eq:m_vir}), by substituting the  $^{12}$CO($J$=3--2) $R_{\rm deconv}$ for the clump radius.

$^{13}$CO($J$=1--0) column densities, $N (^{13}{\rm CO})$ [cm$^{-2}$], are obtained from the following relations under the assumption of local thermodynamic equilibrium:
\begin{eqnarray}
	&&N(^{13}{\rm CO})\;[{\rm cm}^{-2}] \notag \\
	&& \;\;\;\; =2.42 \times 10^{14} 
	\left( \frac{T_{\rm ex}}{1 - {\rm exp}(-5.29/T_{\rm ex}) } \right)
	\left( \frac{\tau (^{13}{\rm CO})}{1-e^{-\tau (^{13}{\rm CO})}} \right) \notag \\
	&& \;\;\;\;\;\;\;\; \left( \frac{1}{f(T_{\rm ex}) - f(T_{\rm bg})} \right)
	\sum T_{\rm mb}(^{13}{\rm CO}) \delta v ,
	\label{eq_N13CO}
\end{eqnarray}
\begin{equation}
	f(T) = \frac{h\nu}{k}~\frac{1}{{\rm exp}(h\nu / kT)-1},
\end{equation}
where $T_{\rm ex}$ is the excitation temperature of $^{13}$CO($J$=1--0), $T_{\rm bg}$ is the brightness temperature of the background radiation ($\sim$2.7 K), $\tau (^{13}{\rm CO})$ is the optical depth and $T_{\rm mb}(^{13}{\rm CO})$ is the observed main beam temperature of the line. 
The optical depth is computed as
\begin{equation}
	\tau (^{13}{\rm CO})
	= - {\rm ln} \left[ 1 - \frac{ T_{\rm mb}(^{13}{\rm CO}) }{ 5.29 [{\rm exp}(5.29/T_{\rm ex}) - 0.164)] }  \right]. 
\label{tau_13co}
\end{equation}
 $\sum T_{\rm mb}(^{13}{\rm CO}) \delta v$ in eq.(\ref{eq_N13CO}) corresponds to the integrated intensity of the $^{13}$CO($J$=1--0) spectrum. Since the beam-filling factors of our targets are unknown, $^{12}$CO($J$=1--0) intensity is useless for estimating the excitation temperature of the clumps. We here supposed $T_{\rm ex} = 10$ K, which corresponds to the excitation temperature of typical molecular clouds in our galaxy \citep[e.g.,][]{Yonekura_etal_1997,Kawamura_etal_1998}.

The column density of molecular hydrogen, $N _{\rm H_2}$ can be obtained from $N(^{13}{\rm CO})$ via the molecular abundance ratios [$^{12}$CO/H$_{2}$] and [$^{12}$CO/$^{13}$CO]. We have used same abundance ratios as the N159 in the LMC \citep{Mizuno_etal_2010}, taking [$^{12}$CO/H$_{2}$] as 1.6 $\times$ 10$^{-5}$ and [$^{12}$CO/$^{13}$CO] as 50. 
Thus, we assumed the following relation between $N _{\rm H_2}$ and $N(^{13}{\rm CO})$:
\begin{equation}
N({\rm H}_{2}) = \frac{50}{1.6 \times 10^{-5}}N(^{13}{\rm CO}).
\label{eq_H2column}
\end{equation}
Finally, the masses $M_{\rm clump}$ of the $^{13}{\rm CO}$ clumps are estimated from 
\begin{equation}
	M_{\rm clump} = \mu m({\rm H}_{2})\sum [D^{2} \Omega N({\rm H}_{2})],
\label{eq_ltemass}
\end{equation}
where $\mu$ is the mean molecular weight per hydrogen molecule, assumed to be 2.8 by taking into account a relative abundance of 71\% hydrogen, 27\% helium, and 2\% metals in mass. $m$(H$_{2}$) is the H$_{2}$ molecular mass, $D$ is the distance to the object (in the case of LMC, $D \sim 50$ kpc), and $\Omega$ is the solid angle subtended by the clumps ($D^2 \Omega \sim \pi R ^{2}_{\rm deconv}$).
We summarize these physical properties in table \ref{tab_13CO}. Mean values in the N48 and N49 regions are $M_{\rm vir} \sim 1.8 \times 10^4 \;{\rm M}_{\odot}$, $N({\rm H_{2}}) \sim 2.3 \times 10^{21}\; {\rm cm}^{-2}$, and $M_{\rm clump} \sim 1.0 \times 10^4 \;{\rm M}_{\odot}$.    

\subsection{Large velocity gradient analysis}
\subsubsection{Temperature and density estimation via LVG modelling}
To estimate the kinetic temperature and number density of the molecular gas, we have performed a large velocity gradient analysis \citep[LVG analysis;][]{Goldreich_Kwan_1974,Scoville_Solomon_1974} of the CO rotational transitions. The LVG radiative transfer code simulates a spherically symmetric cloud of uniform density and temperature with a spherically symmetric velocity gradient proportional to the radius, and employs a Castor's escape probability formalism \citep{Castor_1970}. The LVG model requires three independent parameters to computer emission line intensities: the kinetic temperature, $T_{\rm kin}$, the density of molecular hydrogen, $n({\rm H}_2)$, and $X({\rm CO})/(dv/dr)$. $X({\rm CO})/(dv/dr)$ is the abundance ratio of CO to H$_2$ divided by the velocity gradient in the cloud. We adopt uniform fractional abundance ratios [$^{12}$CO/H$_{2}$] of 1.6 $\times$ 10$^{5}$, [$^{12}$CO/$^{13}$CO] of 50, as in \S \ref{sec_13COclump}. We estimate the mean velocity gradient $dv/dr=\Delta V_{\rm clump}/2 R_{\rm deconv}$, which is the same estimate with \cite{Minamidani_etal_2008}. They checked that the change in $dv/dr$ caused by the definition of clump boundary does not largely affect the LVG calculations.

We performed the LVG model calculations over a grid of temperatures in the range $T_{\rm kin}$ = 5--200 K and densities in the range $n({\rm H}_2)$ = 10--10$^6$ cm$^{-3}$, with grid spacings of $10^{0.02}$ K and $10^{0.125}$ cm$^{-3}$, respectively. This produced sets of line intensity ratios, $R ^{12}_{3-2/1-0}$ and $R^{12/13}_{1-0}$. The model includes the lowest 40 rotational levels of the ground vibrational state and uses the Einstein $A$ and H$_2$ impact rate coefficients of \cite{Schoier_etal_2005}. We do not include higher energy levels, which would require including populations in the lower vibrationally excited states. This imposes a limit of $T_{\rm kin} \sim 200$ K in the present study, and even higher temperatures are not in general excluded below.

To solve for the temperatures and densities that best reproduce the observed line intensity ratio, we calculate chi-squared as
\begin{equation}
	 \chi ^{2} = \sum _{i=1}^{N-1} \sum _{j=i+1}^{N} 
	 	\frac{[R _{\rm obs}(i,j) - R_{\rm LVG}(i,j)]^2 }{\sigma (i,j)}
\end{equation}
where $N$ is the number of transitions of the observed molecule (in this case $N=3$), $i$ and $j$ refer to different molecular transitions, $R_{\rm obs}(i, j)$ is the observed line intensity ratio from transition $i$ to transition $j$, and $R_{\rm LVG}(i, j )$ is the ratio between transitions $i$ and $j$, estimated from the LVG calculations. The standard deviation, $\sigma (i, j )$, for $R_{\rm obs}(i, j )$ is estimated from the noise level of the observations and the calibration uncertainties.

\subsubsection{Results of the LVG analysis \label{lvg}}
We derived kinetic temperatures and number densities for the clumps via the LVG analysis described above. The analysis was performed for clumps covered by the Mopra observations, whose $^{13}$CO($J$=1--0) line intensities $T_{\rm mb}$ are greater than 3 $\sigma$ noise level at the $^{12}$CO($J$=3--2) clump peak positions. The selected clumps are N48--1, 2, 3, 4, 5, 6, 7, 8, 14, N49--1, 2, 3 (Hereafter N4849 clumps). $R ^{12}_{3-2/1-0}$ and $R^{12/13}_{1-0}$ were calculated observationally from the ratios of the peak main beam temperature values, where the $^{12}$CO($J$=3--2) data has been smoothed to the resolution of the Mopra data. The uncertainties of the two ratios are estimated to be 8\%--28\% and 10\%--31\%, respectively, calculated by combining the calibration errors and the formal errors on the gaussian model fits. We also performed the same analysis on the clumps of \cite{Minamidani_etal_2008,Minamidani_etal_2011} (These authors also performed their own LVG analysis, but we repeat the calculations using the method and parameters described above, in order that the results may be consistently compared). The $\chi$-squared method enables us to derive a best solution for the physical parameters, which had sometimes been difficult to estimate from the overlap of two ratios \citep[see Appendix A of ][]{Minamidani_etal_2008}. The selected clumps are 30Dor--1, 3, 4, N159--1, 2, 4, N166--1, 3, 4, N206--1, 2, N206D--1, GMC225--1, 3 \citep[see table 2 of][Hereafter M0811 clumps]{Minamidani_etal_2008}. We summarize the parameters used in the LVG analysis in Table \ref{tab_lvgpara}.

We regarded the lowest values of $\chi ^2$ as the best solution for the temperatures and the densities of the clumps, and define error bars as $\chi ^2 < 3.84$, which corresponds to the 5\% confidence level of the $\chi ^2$ distribution with one degree of freedom. We omitted those clumps whose lowest $\chi ^2$ value is greater than 3.84. When the valid region of parameter space exceeds the boundary of the calculated range, $T_{\rm kin} =$ 5--200 K and $n({\rm H}_2) = $10--10$^6$ cm$^{-3}$, we treat the the edge of the calculated range as the true boundary. Contour plots of LVG analysis on the $n({\rm H}_2)$--$T_{\rm kin}$ plane are summarized in Appendix A (Figures \ref{fig_nTplot_N4849} and \ref{fig_nTplot_M0811}), and derived values are summarized in Table \ref{tab_lvgresults}. 
 
Although N48--4 and 30 Dor--1 are omitted from the calculation due to their high minimum $\chi ^2$ values (exceed 3.84), it is still possible to obtain an estimate of their properties. In \cite{Minamidani_etal_2011}, the LVG results of 30 Dor--1 with different line ratios are quoted as $T_{\rm kin} = 134^{+66}_{-49}$ K and $n({\rm H_2}) = 4.6^{+2.4}_{-1.3} \times 10^3$ cm$^{-3}$, making it one of the densest and warmest clumps in their sample. Since the LVG results of N48--4 are similar to those of 30 Dor--1, and $R^{12}_{3-2/1-0}$ is high in both clumps ($\simeq$1.3), it is possible that the temperatures and densities of N48--4 may be as high as that of 30 Dor--1.

\begin{figure}
\epsscale{1.0}
\plotone{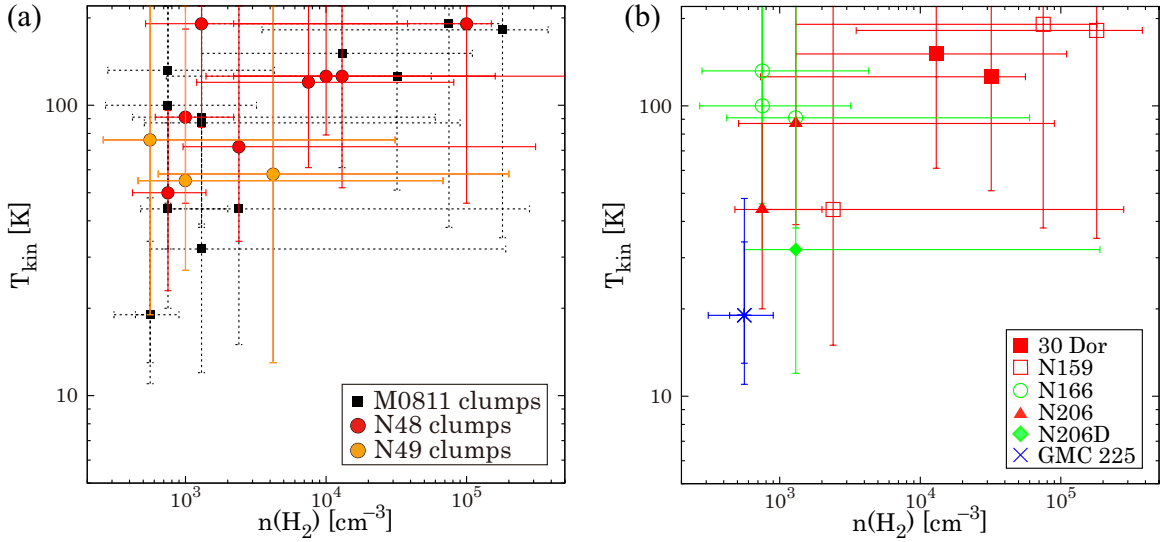}
\caption{(a) Plot of molecular hydrogen density ($n$(H$_2$)) and kinetic temperature ($T_{\rm kin}$) of the clumps derived via LVG analysis. Red and orange filled circles indicate the clumps in the N48/N49 regions (N4849 clumps), and black filled boxes indicate the clumps of \cite{Minamidani_etal_2008,Minamidani_etal_2011} (M0811 clumps). Error bars (solid line for N4849 clumps and dotted line for M0811 clumps) are the range of values for which $\chi ^2$ is less than 3.84. (b) $n$(H$_2$)--$T_{\rm kin}$ plot of the M0811 clumps. Different symbols indicate the various regions in which the clumps are located. Each mark and error bar is colored according to the types of their parental GMCs; Type I in blue, Type II in green, Type III in red.  \label{fig_nTplot}}
\end{figure}

The output of the LVG calculations are shown as $n({\rm H}_2)$--$T_{\rm kin}$ plots in Figure \ref{fig_nTplot}(a)(b). The N4849 clumps are typically warm (greater than $50$ K), with a wide distribution of densities ($10^3$--$10^5$ cm$^{-3}$). $n({\rm H}_2)$ and $T_{\rm kin}$ of the N48 clumps tend to be higher than those of the N49 clumps. In figure \ref{fig_nTplot}(b), the M0811 clumps are colored according to their parent GMC types defined in \cite{Kawamura_etal_2009}: ``starless'' molecular clouds in the sense that they are not associated with \ion{H}{2} regions or young clusters (Type I); molecular clouds with \ion{H}{2} regions (Type II); and molecular clouds with \ion{H}{2} regions and young clusters (Type III). The clumps in active star forming GMCs (Type III: 30 Dor, N159) tend to be warm and dense, and the clumps in starless GMCs (Type I: GMC225) tend to be colder and less dense. These tendencies suggest an evolutionary sequence in terms of increasing density leading to star formation and increased star formation activity leading to intense FUV photons that heat the clouds, as discussed in \cite{Minamidani_etal_2008,Minamidani_etal_2011}. 
The $n({\rm H}_2)$--$T_{\rm kin}$ plots of the N4849 clumps are distributed among those of M0811 clumps hosted by Type II and Type III GMCs. This suggests that the N48 region GMC may be just at the stage of evolving from Type II to Type III. This is consistent with the suggestion of the GMCs in the N48/N49 region are in the early stage of a cluster-forming cloud \citep{Mizuno_etal_2001}. 

We note that here uniform fractional abundance ratios [$^{12}$CO/H$_{2}$] of 1.6 $\times$ 10$^{5}$ have been adopted, which was used in \cite{Mizuno_etal_2010} and \cite{Minamidani_etal_2011}. Different abundance ratios give different solutions (higher temperature and lower density for a higher abundance ratio), but the differences are not significant \citep{Minamidani_etal_2008}. We consider the assumption of uniform fractional abundance to be reasonable, and to not affect the relative tendencies of the results.

\section{Star formation activity and clump properties \label{04} }
	To investigate the relation between star formation activity and the physical parameters of the clumps, we compare the $^{12}$CO($J$ =3--2) distribution and derived clump properties with H$\alpha$ and infrared data.

We use H$\alpha$ data from the Magellanic Cloud Emission-Line Survey \citep{MCELS_1999}, which was obtained with a CCD camera on the Curtis Schmidt Telescope at Cerro Tololo Inter-American Observatory. The images were taken with 2048 $\times$ 2048 CCD cameras that provide an angular resolution of $\sim$ 3$^{\prime \prime}$--4$^{\prime \prime}$ (corresponding to 0.75--1.0 pc at the LMC) and were taken with a narrowband interference filter centered on the H$\alpha$ line ($\lambda _{\rm c}$ = 6563 \AA, $\Delta\lambda $ = 30 \AA).

The LMC has been surveyed by the Spitzer Space Telescope using both the Infrared Array Camera \citep[IRAC;][]{Rieke_etal_2004} and the Multiband Imaging Photometer for Spitzer \citep[MIPS;][]{Fazio_etal_2004} under the Legacy Program ``SAGE'' \citep{Meixner_etal_2006}. The observations were made in the IRAC 3.6, 4.5, 5.8, and 8.0 $\mu$m and MIPS 24, 70, and 160 $\mu$m bands in July and October-November, 2005. In this paper, we use MIPS 24 $\mu$m band and IRAC 8.0 $\mu$m band data. The angular resolutions are $\sim$6$^{\prime \prime}$ for the 24 $\mu$m band and $\sim$2$^{\prime \prime}$ for the 8.0 $\mu$m band, corresponding to 0.5 and 1.5 pc at the distance of the LMC. 
We also use the Spitzer young stellar object (YSO) candidate lists from \cite{Whitney_etal_2008} and \cite{Gruendl_Chu_2009} as a tracer of current star formation activity. The masses of the YSOs are uncertain, but in general those brighter than 8.0 mag at 8.0 $\mu$m represent massive stars with masses greater than $\sim$10 M$_{\odot}$ and those fainter represent intermediate-mass stars with masses of $\sim$4--10 M$_{\odot}$ \citep{Chen_etal_2009}. These YSO lists do not include low-mass YSOs as they cannot be distinguished from background galaxies in the diagnostic color-magnitude diagrams.

\subsection{Star Formation in the N48/49 region}
\begin{figure}
\epsscale{1.0}
\plotone{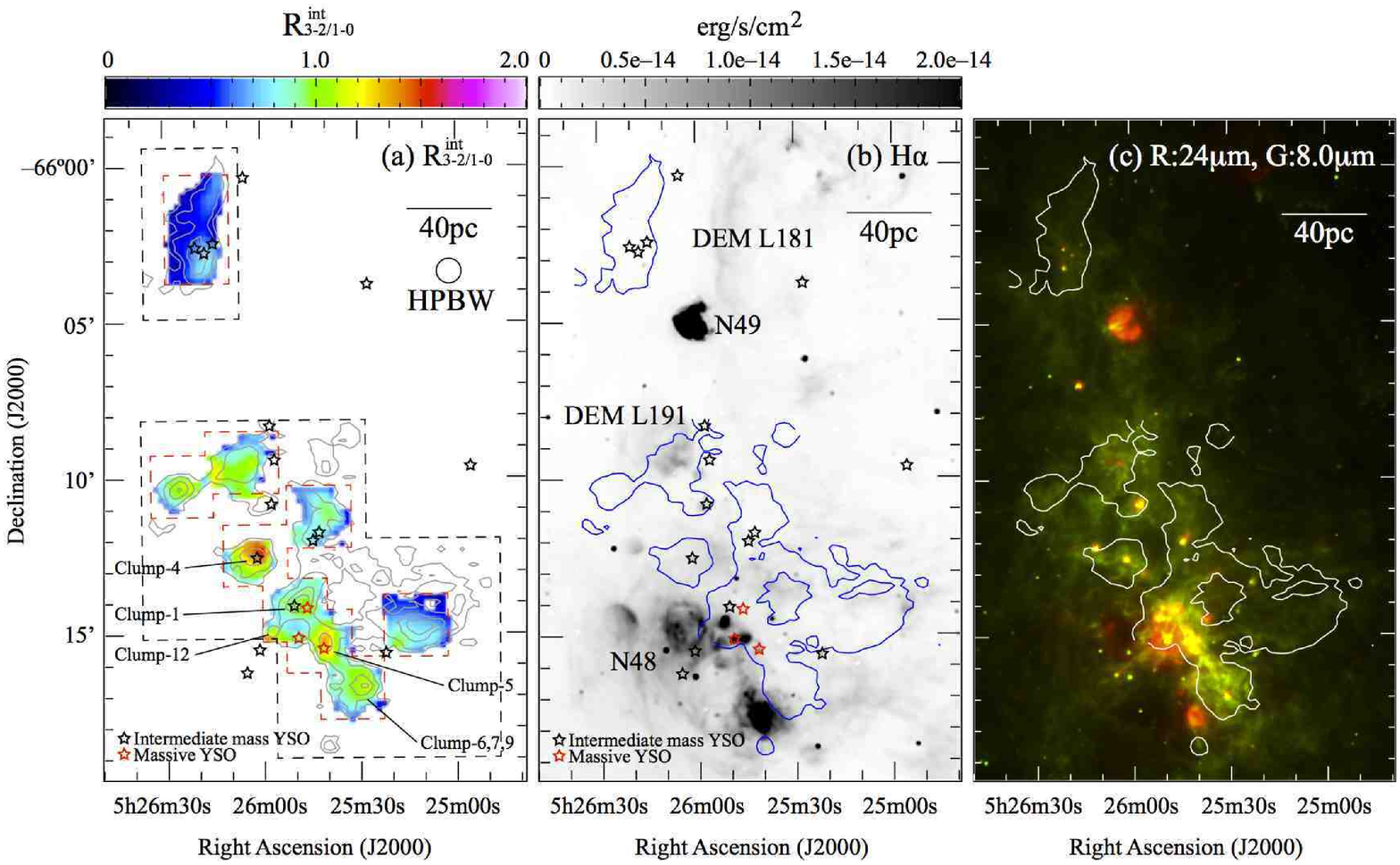}
\caption{Comparison maps of the N48 and N49 regions. Red and black stars in the first two maps are Spitzer YSO candidates \citep{Whitney_etal_2008,Gruendl_Chu_2009}. Red stars indicate massive YSOs ($>$10 M$_{\odot}$), and black stars indicate intermediate mass YSOs (3--10 M$_{\odot}$). 
(a) Color map of the $^{12}$CO($J$=3--2) to $^{12}$CO($J$=1--0) integrated intensity ratio ($R_{3-2/1-0}^{\rm int}$). Black and Red dashed lines correspond to the areas observed with ASTE and Mopra, respectively. Gray contours are the $^{12}$CO($J$=3--2) integrated intensity. The lowest and the second lowest contours are 2.0 and 6.0 K km s$^{-1}$ for the N48 region, and 3.4 and 7.8 K km s$^{-1}$ for the N49 regions, and thereafter run in steps of 8.0 K km s$^{-1}$. (b) Grayscale image of H$\alpha$ flux \citep{MCELS_1999}. Blue contours are the lowest contour of $^{12}$CO($J$=3--2) shown in (a). The names of the H$\alpha$ emission nebulae listed in Table \ref{tab_Hanebula} are also marked. (c) Two color image of Spitzer 24 $\mu$m (red) and 8.0 $\mu$m (green) flux \citep{Meixner_etal_2006}, with the lowest contour of $^{12}$CO($J$=3--2) (white contour) overplotted. \label{com_fig1}}
\end{figure}

Maps of the $^{12}$CO($J$=3--2) to $^{12}$CO($J$=1--0) integrated intensity ratio (hereafter $R_{3-2/1-0}^{\rm int}$) are shown in Figure \ref{com_fig1}(a). Note that the ASTE data is smoothed to the same effective resolution as the Mopra maps (45$^{\prime \prime}$) before this ratio is taken. These are a useful means of showing relationships between the physical conditions of the clumps and the spatial distribution of star formation tracers -- H$\alpha$ nebulae, YSOs the IR dust emission. 
Mean values of $R_{3-2/1-0}^{\rm int}$ are 0.9 for the N48 region and 0.5 for the N49 region, with errors of 0.085 and 0.14, respectively (combination of the noise rms and calibration uncertainties). The clumps in the N48 region show significantly higher $R_{3-2/1-0}^{\rm int}$ than those in the N49 region, even taking into account the relatively large calibration error of the N49 data. In the N48 region, $R_{3-2/1-0}^{\rm int}$ is typically higher than the unity around the peak position of each clump, indicating that the CO molecules are highly excited due to high density or temperature of the clumps.

Figure \ref{com_fig1}(b) shows the H$\alpha$ flux distribution and positions of the Spitzer YSO candidates. \ion{H}{2} regions and an SNR identified by \cite{Henize_1956} and \cite{Davies_Meaburn_1976} are indicated in the figure and also listed in Table \ref{tab_Hanebula}. 
The H$\alpha$ emission nebulae take spatial offset from the clumps, and the clumps have no prominent H$\alpha$ emission inside them. This indicates that prominent \ion{H}{2} regions have not formed yet inside the molecular clumps. 
However, since H$\alpha$ emission is strongly affected by dust extinction, it is possible that \ion{H}{2} regions are embedded within or hidden behind the molecular gas. 
We compared the $^{12}$CO($J$=3--2) distribution with 1.4 GHz radio continuum data \citep{Filipovic_etal_1995,Filipovic_Staveley-Smith_1998,Hughes_etal_2007} and confirm that there is no prominent 1.4 GHz emission where H$\alpha$ emission is absent, and hence that there are no such hidden \ion{H}{2} regions behind or within the molecular clumps.

Figure \ref{com_fig1}(c) shows a 24 and 8.0 $\mu$m 2-color image, with $^{12}$CO($J$=3--2) contours and Spitzer YSO candidates overlaid. These mid-infrared bands are a useful probe of embedded star-formation activity \citep{Calzetti_etal_2007}, since 24 $\mu$m emission is primarily due to thermal emission from hot dust heated by nearby stars, and diffuse 8.0 $\mu$m emission mainly arises from polycyclic aromatic hydrocarbon (PAH) molecules \citep{Li_Draine_2001} which are fluorescing in the local UV radiation in the neutral zones of the photodissociation regions \citep[PDRs; e.g.,][]{Calzetti_etal_2007}. The 24 and 8.0 $\mu$m emission is well correlated with the molecular clumps, especially in the N48 region. A prominent 24 $\mu$m feature is seen in the southern part of the N48 region (around Clump-1 and 5), indicating that the most active star formation is occurring around this area. The 8.0 $\mu$m emission shows quasi-linear bar-like distribution running from north to south. As discussed in \cite{Cohen_etal_2003}, this might be indicative of a region of high-density gas swept up by the expansion of LMC 4 and LMC 5, producing a dense photodissociation region in the dense molecular clouds. 
PAH abundances are believed to be enhanced by ISM shocks through shock fragmentation of graphite grains \citep[e.g.,][]{Jones_etal_1996}. So it is also possible that the PAH molecules in these regions were produced by the large-scale shock of the SGS collision and have survived only in and around the molecular clumps with high enough column densities to shield them from local UV radiation.

\begin{figure}
\epsscale{0.7}
\plotone{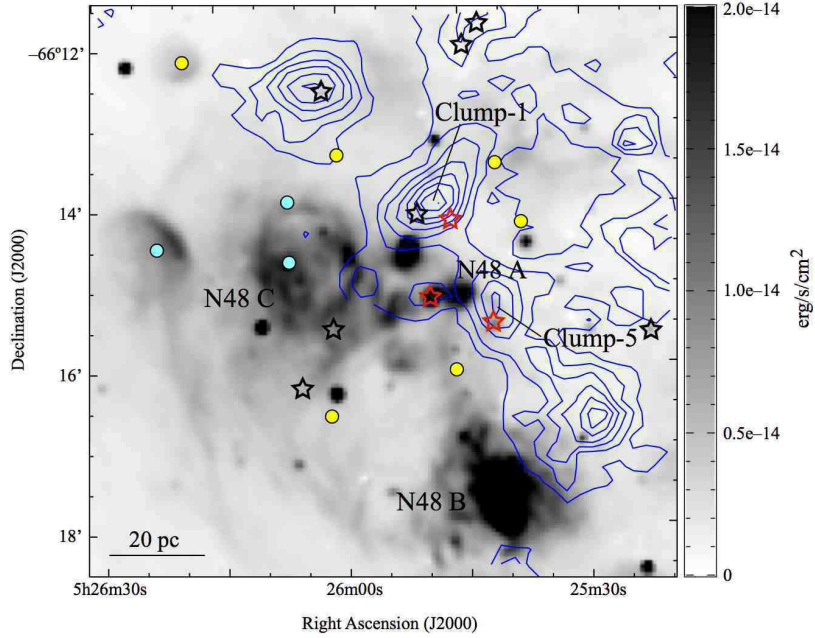}
\caption{Close-up view of the \ion{H}{2} region N48. The grayscale is H$\alpha$ flux and blue contours are $^{12}$CO($J$=3--2). The lowest CO contours correspond to the 5 $\sigma$ noise level (2.0 K km s$^{-1}$ and 3.4 K km s$^{-1}$ for the N48 and N49 regions, respectively) and run in steps of 4.0 K km s$^{-1}$. Red and black stars are Spitzer YSO candidates (same as Figure \ref{com_fig1}), blue and yellow circles are O-type and B-type stars, respectively \citep{Will_etal_1996}. \label{ha_zoom}}
\end{figure}

The YSO candidates are distributed in and around the molecular clumps, suggesting that recent star formation has occurred in both the N48 and N49 regions. Massive YSO candidates are only found in the N48 region, where the prominent 24 $\mu$m emission is located. 
Figure \ref{ha_zoom} shows a zoomed-in view of Figure \ref{com_fig1}(a) around the N48 H$\alpha$ nebulae with additional plots of previously-identified OB stars \citep{Will_etal_1996}. Note that the sample consists of 11 targets that are luminous in V-band (brighter than 15 mag). 
These OB stars are located away from the molecular clumps, which indicates that their parental clouds have already been dispersed by UV radiation or stellar winds. On the other hand the YSO candidates are preferentially located inside the clumps. The ages of these OB stars are estimated to be 5--10 Myr, and the age of the YSOs can be estimated to be $\sim$1 Myr, assuming a YSO formation time scale of 0.2 Myr \citep{Whitney_etal_2008}. These features are well interpreted as a time sequence or age gradient of the star formation activity in this area.

The distribution of $R_{3-2/1-0}^{\rm int}$ shows minimal correlation with that of H$\alpha$ emission. 
For those N48 clumps that are located beside \ion{H}{2} regions (Clump-1, 5, 6, 7, 9, 12), the ratios are relatively high around the center of the clumps and decrease towards the rims of the clumps. An increase of the ratio towards the \ion{H}{2} regions is only seen around Clump-12, the only example in which the effect of UV radiation on the ratio is clearly seen. On the other hand, $R_{3-2/1-0}^{\rm int}$ shows a moderate spatial correlation with the YSOs and the infrared emission, especially the 8.0 $\mu$m flux. These indicate that $R_{3-2/1-0}^{\rm int}$ is more (effectively) enhanced by internal heating sources such as the YSOs and the PDR than by \ion{H}{2} regions.

\subsection{Gravitational state of the clumps and star formation}
Here the gravitational state of the clumps based on the clump mass derived from the $^{13}$CO($J$=1--0) line is discussed. 
A plot of $M _{\rm vir}$ against $M _{\rm clump}$ is displayed in Figure \ref{plot_13vir}, in which the clumps are classified according to their spatial correlation with Spitzer YSO candidates. Here $M _{\rm vir}$ may be thought of as an estimate of the mass required for a clump to be self-gravitating, and $M _{\rm clump}$ as an estimate of the luminosity mass within an assumption of local thermodynamical equilibrium. It is notable that the absolute value of both mass have errors due to the definition of the clump boundary and the estimate of relative molecular abundances, but these errors are less significant when we compare the relative differences between the clumps. The values of $M _{\rm vir}/M _{\rm clump}$ are shown in Table \ref{tab_13CO}. The clumps with YSOs tend to be closer to $M _{\rm vir}/M _{\rm clump} \sim 1$ than the clumps without YSOs, indicating that the clumps with YSOs are relatively gravitationally relaxed than the clumps without YSOs. The best-fit line to the full sample is $M _{\rm vir} = 40 M _{\rm clump}^{0.65}$ with a correlation coefficient 0.68. This corresponds to $M _{\rm vir}/M _{\rm clump} \propto M _{\rm clump}^{-0.35}$. This trend of decreasing $M _{\rm vir}/M _{\rm clump}$ with increasing $M _{\rm clump}$ is also reported in previous Galactic and LMC works (e.g., Orion: \cite{Ikeda_etal_2007, Ikeda_etal_2009}; Cygnus: \cite{Dobashi_etal_1996}; Cepheus and Casseopeia: \cite{Yonekura_etal_1997}; entire the LMC: \cite{Wong_etal_2011}; and 30 Dor: \cite{Indebetouw_etal_2013}).
The power-law index of $-$0.35 for the N48/N49 clumps is smaller than the theoretical value of $-$2/3 for pressure-confined and magnetized clumps \citep[$M _{\rm vir}/M _{\rm clump} \propto M _{\rm clump} ^{-2/3}$;][]{Bertoldi_McKee_1992} and is rather similar to that of Orion ($-$0.33) and Cepheus and Casseopeia ($-$0.38). This suggests that the ratio of the internal kinetic energy to the gravitational energy in the N48 and N49 clumps is relatively low and the N48 and N49 clumps are gravitationally relaxed. 
However, this is a tentative result because the uncertainty of the power-law index is large due both to the small number of data points and to the uncertainties in both masses, mainly arising from the estimates of clump radius. There is no significant difference in the masses of the clumps with massive YSOs and those with intermediate mass YSOs. Although observations at higher spatial resolution ($\sim$2 pc) with higher density tracers (HCO$^{+}$ and HCN) in the LMC have revealed that massive YSO bearing clumps tend to be larger and more massive than those without signs of star formation \citep{Seale_etal_2012}, the correlation between the mass of the $^{13}$CO clumps and the YSOs cannot be seen in our data at a spatial resolution of 11 pc. 
Note that here we have assumed a uniform excitation temperature of 10 K in the derivation of $M_{\rm clump}$, and this might be a lower limit for the N48/N49 region since it corresponds to typical excitation temperatures of Galactic molecular clouds with fewer signs of star formation \citep[e.g., the Gemini and Auriga regions;][]{Kawamura_etal_1998}. 
However, this does not significantly affect the trends seen in the mass ratio, for instance, if we assume $T_{\rm ex}=20$ K, $M _{\rm clump}$ increases by only a factor of $\sim 1.4$.

\subsection{Correlations between $n({\rm H}_2)$, $T_{\rm kin}$ and star formation activity \label{lvg_sf}}
\begin{figure}
\epsscale{1.0}
\plotone{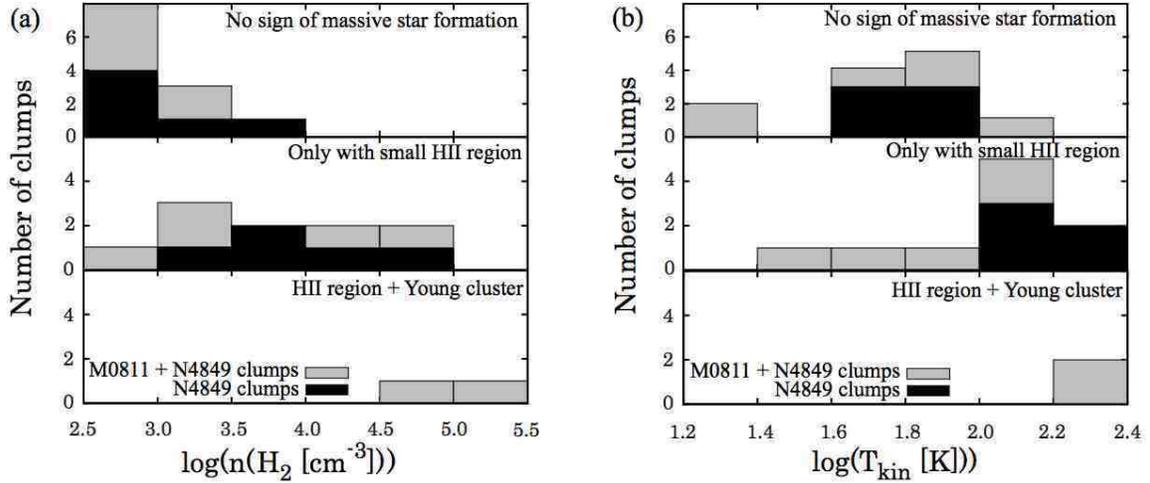}
\caption{Histograms of $n$(H$_2$) and $T_{\rm kin}$ for the N4849 clumps and M0811 clumps, separated by clump type. Top: Clumps without associated O stars capable of ionizing an \ion{H}{2} region; Middle: Clumps with small \ion{H}{2} regions; Bottom: Clumps with stellar clusters and \ion{H}{2} regions. Black boxes are the N4849 clumps and gray boxes are the total sample of clumps. \label{hist_LVG}}
\end{figure}

\cite{Minamidani_etal_2008, Minamidani_etal_2011} discussed the relationship between $n({\rm H}_2)$ and $T_{\rm kin}$ in molecular clumps with their parent GMC types, as defined by \cite{Kawamura_etal_2009} (see \S \ref{lvg}). However, GMC types are defined by star formation activity on the size scale of entire GMCs, which have typical radii of 30--50 pc \citep{Fukui_etal_2008}. Here we instead assign as similar ``type'' to the N4849 and M0811 clumps based on the star formation activity in and around them on $\sim$10 pc size scales.
We sort the clumps into three caegories: Type I: Clumps without O stars capable of ionizing an \ion{H}{2} region, which does not exclude the possibility of associated young stars later than B type; Type II: Clumps with small \ion{H}{2} regions only; Type III: Clumps with stellar clusters and \ion{H}{2} regions. We used the catalog of \cite{Henize_1956} and \cite{Davies_Meaburn_1976} for \ion{H}{2} regions and require an overlap of a clump with diffuse H$\alpha$ flux from these regions in order for it to be assigned to Type II or Type III. We used the catalog of \cite{Bica_etal_1996} for young clusters ($<$10 Myr; SWB 0). Figure \ref{hist_LVG} shows the distribution of $n({\rm H}_2)$ and $T_{\rm kin}$ sorted by the clump type. Type I clumps tend to be cooler and less dense, whereas clumps with star formation activity tend to be warmer and denser. In particular, only dense, hot clumps are identified as young star cluster-forming clumps (Type III). This is demonstrates that $n({\rm H}_2)$ and $T_{\rm kin}$ in molecular clumps are related to the star formation activity within them.

\begin{figure}
\epsscale{1.0}
\plotone{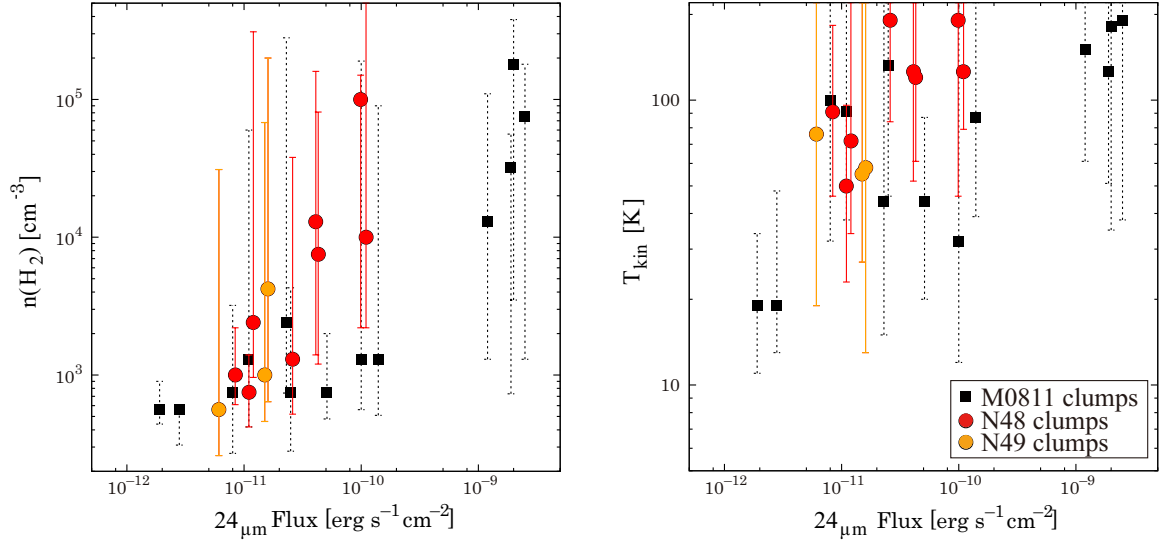}
\caption{Plot of $n$(H$_2$), $T_{\rm kin}$ versus 24 $\mu$m flux density towards the peak positions of the clumps. The symbols are the same as Figure \ref{fig_nTplot}(a). The calibration uncertainty for the 24 $\mu$m flux is typically about two percent, so the error bars are smaller than the size of the plot symbols. \label{plot_LVG_24_nT}}
\end{figure}

We also compare $n({\rm H}_2)$ and $T_{\rm kin}$ with the integrated 24 $\mu$m flux contained within the half power width of the 45$^{\prime \prime}$ beam towards each peak position (Figure \ref{plot_LVG_24_nT}). The derived values are listed in Table \ref{tab_lvgresults}. As displayed in Figure \ref{plot_LVG_24_nT}, 24 $\mu$m flux shows a positive correlation with $n({\rm H}_2)$ and $T_{\rm kin}$. Emission at 24 $\mu$m originates from small dust grains mainly heated by UV photons from young stars, and it has been shown to directly relate to ongoing star formation over a timescale of $\tau _{24} \sim 10$ Myr \citep{Calzetti_etal_2005,Perez-Gonzalez_etal_2006,Calzetti_etal_2007}.
The positive correlation with $n({\rm H}_2)$ indicates that star formation activity is higher in denser molecular clumps, and the positive correlation with $T_{\rm kin}$ indicates that the temperatures of the clumps are higher where the UV radiation is more intense. These positive correlations, including the trend for increased temperature and density with higher levels of star formation activity (Fig. \ref{hist_LVG}), suggest that $n({\rm H}_2)$ and $T_{\rm kin}$ trace an evolutionary sequence of the clumps. Compared with the M0811 clumps, the N4849 clumps, particularly those in the N48, show high densities and temperatures even though 24 $\mu$m flux is not prominent. If density is indeed an effective probe of molecular clump evolutionary state, then these high densities with low 24 $\mu$m flux indicate that the clumps are about to form stars, and star formation will be more active in the near future. High temperatures with low 24 $\mu$m flux suggest that the N4849 clumps are heated effectively due to the lower mass and smaller size, as seen in \S \ref{Clump_def}. There may also be an alternative heating source, potentially something which can not be traced by 24 $\mu$m emission, such as low mass stars.

\section{Discussion \label{05}}
	\subsection{Distribution of \ion{H}{1} and the Molecular Clumps \label{comp_HI}}
\ion{H}{1} gas is diffuse and extended compared with the molecular clouds traced by CO, and it is the better tracer of the global kinematics and structure of the SGSs. The \ion{H}{1} gas distribution therefore allows us to examine the relation between the physical properties of the clumps and the structure of the SGSs. We use \ion{H}{1} survey data taken with the Australia Telescope Compact Array \citep[ATCA;][]{Kim_etal_1998} and the Parkes single-dish telescope \citep{Staveley-Smith_etal_2003}, combined by an image feathering (linear merging) approach \citep{Kim_etal_2003}. The data cube covers the entire LMC at a spatial resolution of 1$^{\prime}$ and a pixel size of 20$^{\prime \prime}$, and has a 1$\sigma$ noise level of 2.4 K in a 1.65 km s$^{-1}$  velocity channel.

\begin{figure}
\epsscale{1.0}
\plotone{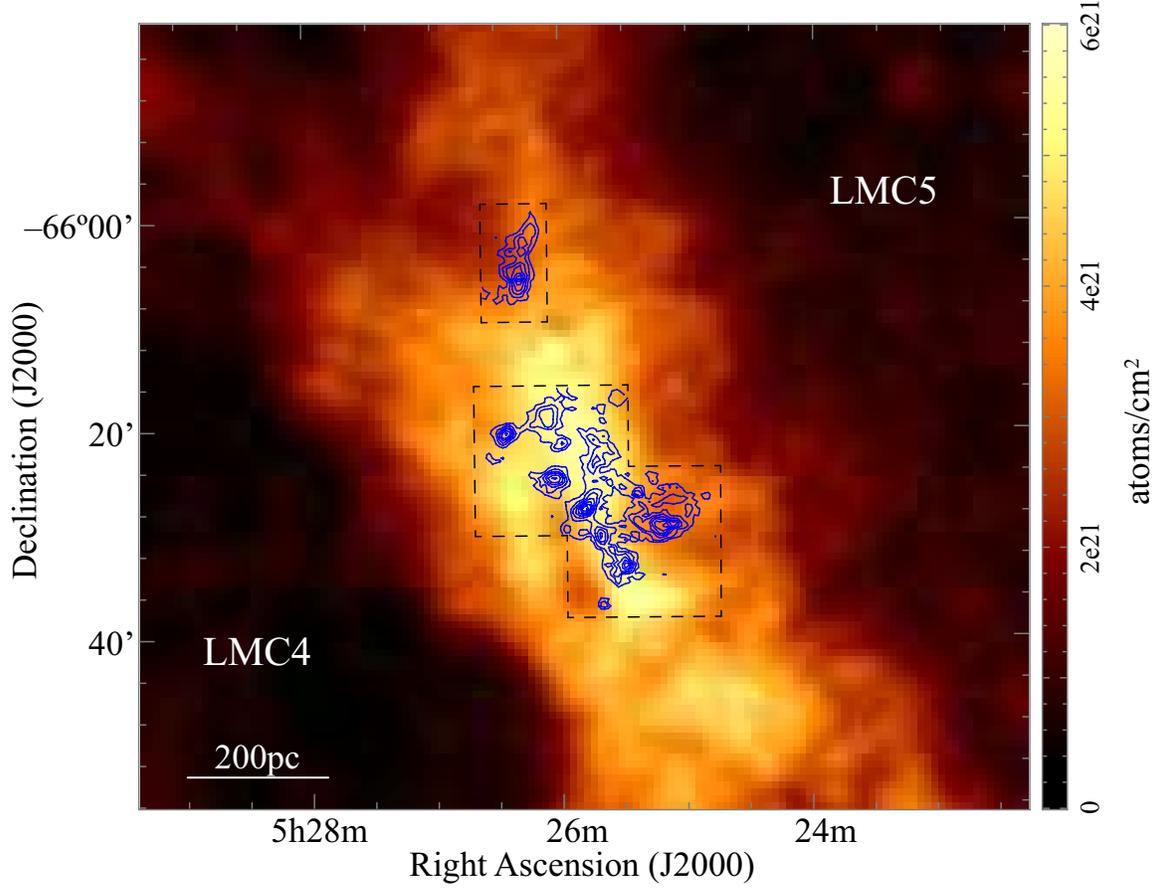}
\caption{$^{12}$CO($J$=3--2) intensity (blue contours) overlaid on an \ion{H}{1} column density map \citep[color scale:][]{Kim_etal_1998,Kim_etal_2003}. The beam sizes of the $^{12}$CO($J$=3--2) and \ion{H}{1} data are 27$^{\prime \prime}$ and 60$^{\prime \prime}$, respectively. The lowest CO contours correspond to the 5 $\sigma$ noise level (2.0 K km s$^{-1}$ and 3.4 K km s$^{-1}$ for the N48 and N49 regions, respectively) and run in steps of 4.0 K km s$^{-1}$. Black dashed lines indicate the area observed with ASTE. \label{fig_HImap}}
\end{figure}

\begin{figure}
\epsscale{1.0}
\plotone{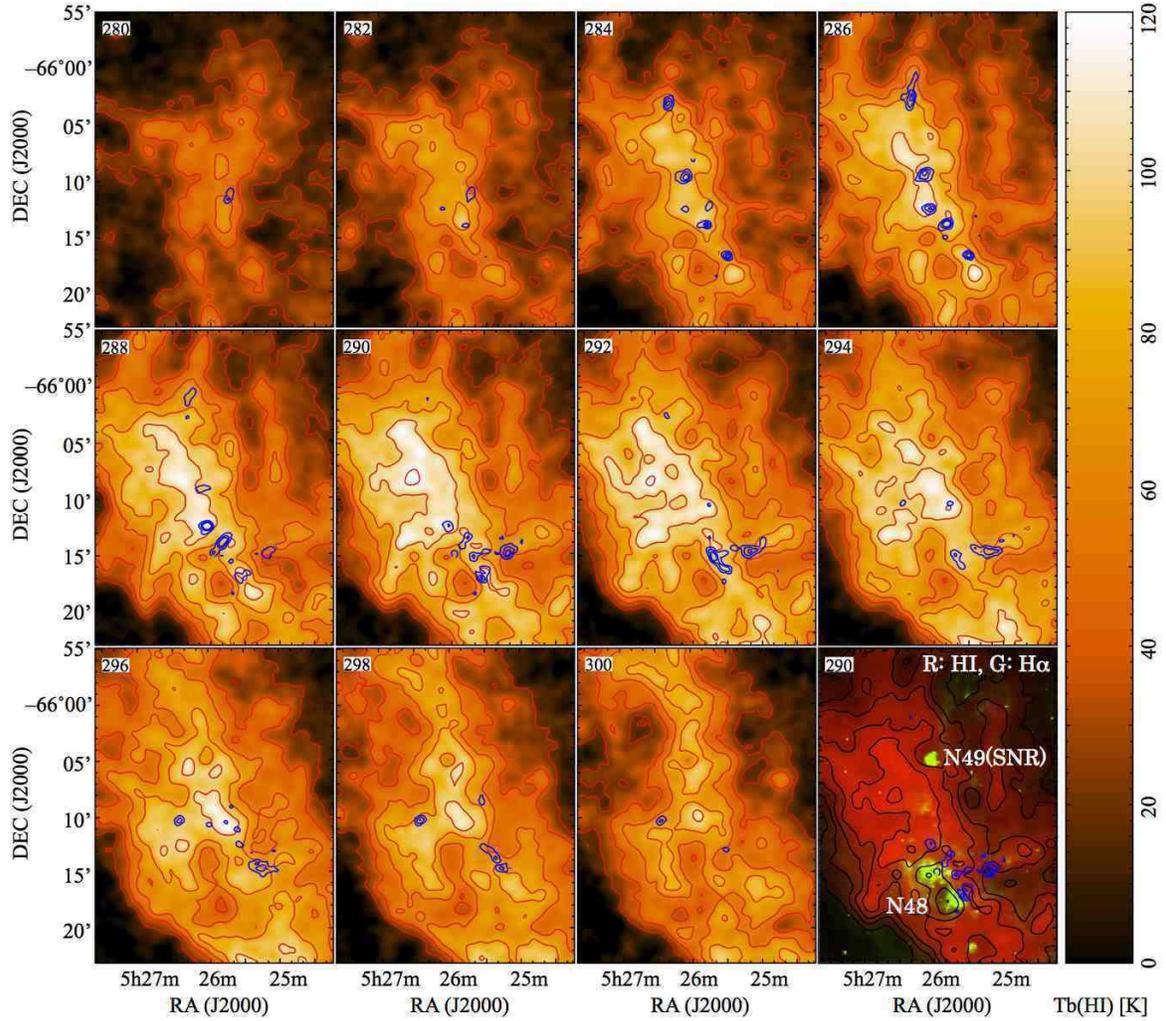}
\caption{Velocity channel maps of the \ion{H}{1} brightness temperature (color image and red contours) and $^{12}$CO($J$=3--2) intensity (blue contour) in the N48/N49 region, averaged in velocity intervals of 2.0 km s$^{-1}$. The central velocity is shown in the upper left of each panel. The beam sizes of the $^{12}$CO($J$=3--2) and \ion{H}{1} data are 27$^{\prime \prime}$ and 60$^{\prime \prime}$, respectively. CO contour levels are from 0.8 (N48) and 1.8 (N49) K km s$^{-1}$ (3$\sigma$) at 1.0 K km s$^{-1}$ intervals. \ion{H}{1} contour levels are from 20 K with 20 K intervals. The bottom right panel shows a two color image of \ion{H}{1} (red) and MCELS H$\alpha$ (green) with \ion{H}{1} contours (black) and CO contours (blue) of the 290 km s$^{-1}$ channel. \label{com_fig4}}
\end{figure}

Figure \ref{fig_HImap} shows a color image of the \ion{H}{1} column density with $^{12}$CO($J$=3--2) contours overlaid. The \ion{H}{1} forms a large ridge, which extends northeast to southwest for several hundred parsecs. The SGSs can be seen as giant cavities in the \ion{H}{1} gas (east: LMC 4, west: LMC 5), and the ridge is located between the two shells. The molecular clumps are distributed along the ridge, and the \ion{H}{1} column density is high (up to $N({\rm H}_2) \sim 6 \times 10^{21} $ cm$^{-2}$) around the N48 clumps.
Figure \ref{com_fig4} shows velocity channel maps of \ion{H}{1} brightness temperature and $^{12}$CO($J$=3--2) intensity. The two tracers show good spatial correlation and the clumps are distributed along the \ion{H}{1} ridge. Then the spatial coincidences in the N48/N49 regions on the size-scales of clumps suggest that the clumps are associated with the \ion{H}{1} envelopes and are formed in the high column density \ion{H}{1} ridge.

There are some offsets in CO and \ion{H}{1} peak positions, particularly obvious in the velocity range 288--292 km s$^{-1}$. These offsets are not surprising, and it has been pointed out that in general \ion{H}{1} envelopes tend to be not perfectly correlated with CO emission, but show some offset in their peak positions, even on the size-scales of GMCs \citep{Fukui_etal_2009}. This can be interpreted as the conversion of \ion{H}{1} into molecular H$_2$ such that the amount of \ion{H}{1} gas is reduced around the molecular clumps. 
Another interpretation is that the \ion{H}{1} envelopes around CO clouds are likely to be the dense and cold atomic gas from which molecular clumps are formed \citep{Fukui_etal_2009}. Such cold, optically thick \ion{H}{1} typically makes only a small contribution to \ion{H}{1} emission profiles, 
meaning that the apparent \ion{H}{1} intensity may be relatively small at the velocities at which these components are present. 
It is likely that both explanations are partially responsible for the observed offsets.
One-dimensional numerical simulations of the propagation of a shock wave into the atomic ISM find that thermal instabilities occur in the shock-compressed layer, which produces geometrically thin, dense layers of gas in which a large amount of hydrogen molecules are formed \citep[e.g.,][]{Koyama_Inutsuka_2000,Hosokawa_Inutsuka_2007}. 2D models of colliding supershells whose terminal sizes reach several hundreds of parsecs \citep{Ntormousi_etal_2011}, suggest that small (typical sizes of $<$ 1 pc), dense (up to 10$^4$ cm$^{-3}$, fulfilling the density requirements for molecule formation), cold (T $<$ 100 K) gas clumps and filaments form naturally by thermal instabilities in the highly turbulent and compressed collisional area of the two supershells. 
It has also been pointed out from observational studies that the cool \ion{H}{1} gas in the LMC is unusually abundant compared to the cold atomic phase of the Milky Way \citep{Dickey_etal_1994,MarxZimmer_etal_2000}. The spatial offset between CO and \ion{H}{1} peaks in the N48/N49 region is consistent with a scenario in which the molecular gas was formed in cold neutral medium that embedded in more extended warm neutral medium. 
The current spatial resolution of \ion{H}{1} data is not high enough for the investigation of the existence of such a small and cold gas, and higher resolution observations will be necessary in order to discuss it further.


Note that there are shell-like \ion{H}{1} holes where the \ion{H}{2} regions N48-B, C, and the SNR N49 are located (Figure \ref{com_fig4}). The spatial agreement of the holes and the H$\alpha$ nebulae can be clearly seen in the bottom right panel of Figure \ref{com_fig4}. In these holes, \ion{H}{1} gas has presumably been dispersed or completely ionized by the UV radiation and the SNR shock. Around the N48 \ion{H}{2} regions, 8.0 $\mu$m flux is also suppressed (Fig. \ref{com_fig1}(d)), indicating that PAH molecules are have also been dispersed or destroyed. These tendencies strongly suggest that the N48 \ion{H}{2} regions and the N49 SNR form local shells, which may be currently expanding (although the spatial resolution of the \ion{H}{1} data is insufficient for expansion to be confirmed). These small shells may also have triggered star formation around them. It is notable that massive YSOs are only found in the clumps around the shells (Figure \ref{com_fig1}), as is the bright 24 $\mu$m point-like feature, which indicates young star formation. The clumps next to these holes (Clump-1, 4, 5, 6, 8, and 9) show higher $n({\rm H}_2)$ and $T_{\rm kin}$ than those isolated from the \ion{H}{2} regions (Table \ref{tab_lvgresults}, Figure \ref{hist_LVG}). However, further investigation is necessary to confirm that these are signs of heating and/or compression of molecular clumps by the \ion{H}{2} regions, because $R_{3-2/1-0}^{\rm int}$ in these clumps does not appear to be affected by their presence (Fig. \ref{com_fig1}(a)).

\subsection{Relation Between the Molecular Clump Properties and their Location with respect to the SGSs}
\begin{figure}
\epsscale{0.6}
\plotone{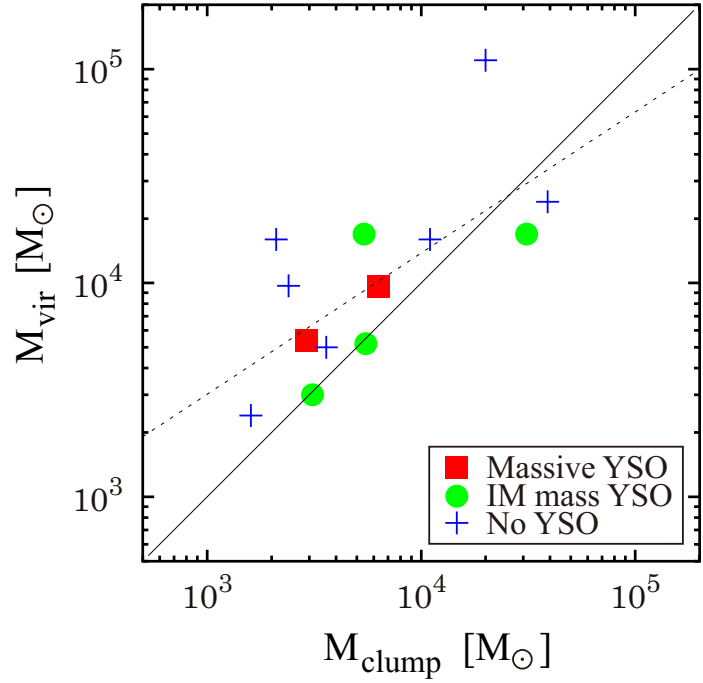}
\caption{Plot of $M _{\rm vir}$ against $M _{\rm clump}$ for the clumps in the N48/N49 region. Red filled boxes and green filled circles indicate clumps associated with massive YSOs and intermediate mass YSOs, respectively, and blue crosses are clumps with no YSO candidates. The solid line represents $M _{\rm vir} = M _{\rm clump}$, and the dashed line is the best-fit line $M _{\rm vir} = 40 M _{\rm clump}^{0.65}$. \label{plot_13vir}}
\end{figure}

The HI distribution suggests that the N49 clumps are associated with only one SGS, LMC5, whereas the N48 clumps are located right in the boundary of the two SGSs. Comparisons of these two regions may therefore provide insight into the differences between the physical effects of a single shell and the interaction of two SGSs on the molecular clump properties and star formation activity.
The excitation state of the molecular clumps, as traced by $R_{3-2/1-0}$ and $R_{3-2/1-0}^{\rm int}$, is apparently higher for the entire population of molecular clumps in the N48 region (Fig. \ref{com_fig1}(b)). The value is typically higher than unity around the clump peak positions, which indicates that the CO molecules are efficiently excited in high density or temperature conditions. $n({\rm H}_2)$ and $T_{\rm kin}$ are higher in the N48 region, and some of the clumps show high densities that are comparable to massive cluster-forming clumps in the LMC molecular ridge and CO arc (Fig. \ref{fig_nTplot}(a)). Although there is no apparent difference in virial state between the two regions (Fig. \ref{plot_13vir}), the dense, warm, and highly excited states of the N48 clumps indicate that they have the more suitable conditions for the formation of massive stars or clusters.
YSO candidates are found in both regions, but massive star formation is more active in the N48 region. There is no sign of massive star formation in the N49 clumps, whereas the N48 region contains several \ion{H}{2} regions, and massive YSOs and luminous 24 $\mu$m and 8.0 $\mu$m emission can be seen in the molecular clumps. The star formation in the N48 region seems to be more evolved compared to the N49 region. These differences of molecular clump properties and star formation evolutionary stage suggest that formation of dense molecular clumps and massive stars is enhanced in the region of high-density gas swept up by the two SGSs, and this SGS interaction has worked more efficiently to create stars from the ISM than the effect of only one SGS.  

\subsection{Evolutionary scenario for the N48/N49 region}
In the previous subsection, we argued that the high density molecular clumps have been formed at the interface of the two SGSs. The spatial correlation between \ion{H}{1} and CO implies that the dense clumps are formed in the high column density \ion{H}{1} ridge, but to conclude that the shells actually trigger formation of these clumps, we must check the consistency of the shell and GMC formation timescales. The dynamical ages of LMC 4 and LMC 5 have been estimated as 4--15 Myr, and 5--7 Myr, respectively, using the current observed expansion velocities of the HI gas \citep{Dopita_etal_1985,Kim_etal_1999,Book_etal_2008}. However, these dynamical ages give roughly lower limits for the shell formation timescales, because they are derived under the assumption that the SGS is formed as a single expanding shell by feedback from a single generation of star formation. 
There is alternative theory that large SGSs might be formed by several generations of star formation \citep{Efremov_Elmegreen_1998}, and age estimates based on stellar population studies in individual SGSs have indicated ages of typically 10--20 Myr \citep[e.g.,][]{Dopita_etal_1985,Points_etal_1999,Glatt_etal_2010}. Particularly in LMC 4, there is an extended stellar arc ($\sim$600pc) in Constellation 3, which is located in the central area of the HI hole, and it has been argued that this arc was formed in the gas swept up by the first generation stellar feedback within the triggering time scale of $\sim$14 Myr \citep{Efremov_Elmegreen_1998}. The arc itself is also an energy source via second generation stellar feedback. We here adopt ages of 10--20 Myr for the two SGSs, with a particularly strong constraint of $>$15 Myr implied for LMC 4. 
Typical timescales for the formation of GMCs from diffuse HI gas have been roughly estimated as $\sim$10 Myr in the LMC \citep{Fukui_etal_2009}, which corresponds to the crossing time of a typical GMC of $\sim$100 pc size assuming a typical \ion{H}{1} velocity dispersion of $\sim$10 km s$^{-1}$ \citep{Fukui_Kawamura_2010}. The free-fall time of \ion{H}{1} gas whose density is $\sim 10$ cm$^{-3}$ is also $\sim$10 Myr. The GMCs in the N48/N49 region are classified as Type II GMCs, which are roughly estimated to have ages of between 6--19 Myr \citep{Kawamura_etal_2009}. These timescales are on the same order of the estimated SGS age of 10--20 Myr, and are consistent with a scenario in which the GMCs in the N48/N49 region were formed by the SGSs.

The suggested evolutionary scenario for the N48 and N49 region can be summarized as follows: 
Dense, young, clumpy, star-forming molecular clumps were formed in the large scale \ion{H}{1}  ridge accumulated by the two SGSs, LMC 4 and LMC 5. The densest molecular clumps ($\sim$10$^4$ cm$^{-3}$), in which star formation is more advanced, were formed at the interaction zone of the two SGSs. Some massive stars have already dispersed their parental molecular gas and formed \ion{H}{2} regions. 
Thus, the large-scale structure of the SGSs affects the formation of the young, star-forming molecular clumps; in particular the interaction of the two SGSs plays an important role and triggers the active formation of dense molecular clumps and massive stars.

\section{Summary}
	We have studied the effects of supergiant shells on dense molecular clumps via $^{12}$CO ($J$=3--2, 1--0) and $^{13}$CO($J$=1--0) observations of the N48/N49 region in the LMC. The main results are summarized as follows:
\begin{enumerate}
\item Our new $^{12}$CO ($J$=3--2) observations, at a spatial resolution of 7pc, have revealed that the GMCs in the N48/N49 region shows highly clumpy structure. We have identified 18 distinct clumps in the N48 region, and 3 clumps in the N49 region. Mean values of the clump physical parameters are: clump size $R_{\rm deconv} \sim 4.7$ pc, line width $\Delta V_{\rm clump} \sim 4.3$ km s$^{-1}$, virial mass $M_{\rm vir} \sim 2.0 \times 10^{4}$ M$_{\odot}$, $^{12}$CO($J$=3--2) luminosity $L_{{\rm CO}(J=3-2)}\sim 1.2 \times 10^3$ K km s$^{-1}$ pc$^2$, respectively. These values are smaller than those of clumps observed in the LMC molecular ridge and the CO Arc with the same instrument \citep{Minamidani_etal_2008}.
\item We have performed LVG radiative transfer calculations in order to estimate the density and temperature of 12 clumps in the N48/N49 region, as well as 14 clumps in the LMC molecular ridge and the CO Arc, using the line intensity rations $R ^{12}_{3-2/1-0}$ and $R^{12/13}_{1-0}$. The N4849 clumps are typically warm ($T_{\rm kin} > 50$ K), with a wide range of number densities ($10^3$--$10^5$ cm$^{-3}$), which indicate that the clumps are in the early stage of star formation. Some N48 clumps are as dense as the massive cluster forming clumps in the LMC molecular ridge region, such as 30 Dor and N159.
\item The ratio of the internal kinetic energy to the gravitational energy is relatively low in the clumps associated with YSO candidates compared to the clumps without YSO candidates, indicating that the the clumps with YSOs are relatively gravitationally relaxed. The mass of the clumps is not correlated with the mass of the associated YSOs.
\item Both the location of molecular clumps in the \ion{H}{1} envelope, and the formation time scales of the SGSs and the GMCs, are consistent with the scenario that the molecular clumps in the N48/N49 region are formed in the high column density \ion{H}{1} ridge in the interface of the two SGSs. 
\item The N48 region is located right at the interaction zone of the two SGSs, whereas N49 is associated with LMC 5 alone. The clumps in the N48 region are typically denser and warmer than those in the N49 region, and the star formation activity, as traced by H$\alpha$ and IR emission, seems to be more evolved. This suggests that the formation of massive clumps and stars proceeds more efficiently where the two SGSs are interacting.

\end{enumerate}

\acknowledgments

We thank the referee, whose thorough reading of the manuscript. K. Fujii thanks Shinya Komugi and Remy Indebetouw for fruitful discussions.
A part of this study was financially supported by MEXT Grants-in-Aid for Scientific Research (KAKENHI) Grant Numbers 15071202, 15071203, 20001003, and JSPS KAKENHI Grant Numbers 14J11419, 22740127.
The ASTE project is managed by Nobeyama Radio Observatory (NRO), a branch of the National Astronomical Observatory of Japan (NAOJ), in collaboration with University of Chile, and Japanese institutes including University of Tokyo, Nagoya University, Osaka Prefecture University, Ibaraki University, and Hokkaido University. Observations with ASTE were carried out remotely from Japan by using NTT's GEMnet2 and its partner R\&E (Research \& Education) networks, which are based on the AccessNova collaboration of University of Chile, NTT Laboratories, and NAOJ.
The 
Mopra Telescope is part of the Australia Telescope which is funded by the Commonwealth of Australia for operation as a National Facility managed by CSIRO.
Cerro Tololo Inter-American Observatory (CTIO) is operated by the Association of Universities for Research in Astronomy Inc. (AURA), under a cooperative agreement with the National Science Foundation (NSF) as part of the National Optical Astronomy Observatories (NOAO). The MCELS is funded through the support of the Dean B. McLaughlin fund at the University of Michigan and through NSF grant 9540747.
SAGE research has been funded by NASA/Spitzer grant 1275598 and NASA NAG5-12595.

\bibliographystyle{yahapj}
\bibliography{thesis}

\begin{thebibliography}{89}
\providecommand\natexlab[1]{#1}
\providecommand\JournalTitle[1]{#1}

\bibitem[{{Bertoldi} \& {McKee}(1992)}]{Bertoldi_McKee_1992}
{Bertoldi}, F., \& {McKee}, C.~F. 1992,
  \href{http://dx.doi.org/10.1086/171638}{\JournalTitle{\apj}, 395, 140}

\bibitem[{{Bica} {et~al.}(1996){Bica}, {Claria}, {Dottori}, {Santos}, \&
  {Piatti}}]{Bica_etal_1996}
{Bica}, E., {Claria}, J.~J., {Dottori}, H., {Santos}, Jr., J.~F.~C., \&
  {Piatti}, A.~E. 1996,
  \href{http://dx.doi.org/10.1086/192251}{\JournalTitle{\apjs}, 102, 57}

\bibitem[{{Book} {et~al.}(2008){Book}, {Chu}, \& {Gruendl}}]{Book_etal_2008}
{Book}, L.~G., {Chu}, Y.-H., \& {Gruendl}, R.~A. 2008,
  \href{http://dx.doi.org/10.1086/523897}{\JournalTitle{\apjs}, 175, 165}

\bibitem[{{Book} {et~al.}(2009){Book}, {Chu}, {Gruendl}, \&
  {Fukui}}]{Book_etal_2009}
{Book}, L.~G., {Chu}, Y.-H., {Gruendl}, R.~A., \& {Fukui}, Y. 2009,
  \href{http://dx.doi.org/10.1088/0004-6256/137/3/3599}{\JournalTitle{\aj},
  137, 3599}

\bibitem[{{Braun} {et~al.}(1997){Braun}, {Bomans}, {Will}, \& {de
  Boer}}]{Braun_etal_1997}
{Braun}, J.~M., {Bomans}, D.~J., {Will}, J.-M., \& {de Boer}, K.~S. 1997,
  \JournalTitle{\aap}, 328, 167

\bibitem[{{Braun} {et~al.}(2000){Braun}, {de Boer}, \&
  {Altmann}}]{Braun_etal_2000}
{Braun}, J.~M., {de Boer}, K.~S., \& {Altmann}, M. 2000, \JournalTitle{ArXiv
  Astrophysics e-prints},
  \href{http://arxiv.org/abs/arXiv:astro-ph/0006060}{{\sffamily
  arXiv:astro-ph/0006060}}

\bibitem[{{Calzetti} {et~al.}(2005){Calzetti}, {Kennicutt}, {Bianchi},
  {Thilker}, {Dale}, {Engelbracht}, {Leitherer}, {Meyer}, {Sosey}, {Mutchler},
  {Regan}, {Thornley}, {Armus}, {Bendo}, {Boissier}, {Boselli}, {Draine},
  {Gordon}, {Helou}, {Hollenbach}, {Kewley}, {Madore}, {Martin}, {Murphy},
  {Rieke}, {Rieke}, {Roussel}, {Sheth}, {Smith}, {Walter}, {White}, {Yi},
  {Scoville}, {Polletta}, \& {Lindler}}]{Calzetti_etal_2005}
{Calzetti}, D., {Kennicutt}, Jr., R.~C., {Bianchi}, L., {et~al.} 2005,
  \href{http://dx.doi.org/10.1086/466518}{\JournalTitle{\apj}, 633, 871}

\bibitem[{{Calzetti} {et~al.}(2007){Calzetti}, {Kennicutt}, {Engelbracht},
  {Leitherer}, {Draine}, {Kewley}, {Moustakas}, {Sosey}, {Dale}, {Gordon},
  {Helou}, {Hollenbach}, {Armus}, {Bendo}, {Bot}, {Buckalew}, {Jarrett}, {Li},
  {Meyer}, {Murphy}, {Prescott}, {Regan}, {Rieke}, {Roussel}, {Sheth}, {Smith},
  {Thornley}, \& {Walter}}]{Calzetti_etal_2007}
{Calzetti}, D., {Kennicutt}, R.~C., {Engelbracht}, C.~W., {et~al.} 2007,
  \href{http://dx.doi.org/10.1086/520082}{\JournalTitle{\apj}, 666, 870}

\bibitem[{{Castor}(1970)}]{Castor_1970}
{Castor}, J.~I. 1970, \JournalTitle{\mnras}, 149, 111

\bibitem[{{Chen} {et~al.}(2009){Chen}, {Chu}, {Gruendl}, {Gordon}, \&
  {Heitsch}}]{Chen_etal_2009}
{Chen}, C.-H.~R., {Chu}, Y.-H., {Gruendl}, R.~A., {Gordon}, K.~D., \&
  {Heitsch}, F. 2009,
  \href{http://dx.doi.org/10.1088/0004-637X/695/1/511}{\JournalTitle{\apj},
  695, 511}

\bibitem[{{Chernin} {et~al.}(1995){Chernin}, {Efremov}, \&
  {Voinovich}}]{Chernin_etal_1995}
{Chernin}, A.~D., {Efremov}, Y.~N., \& {Voinovich}, P.~A. 1995,
  \JournalTitle{\mnras}, 275, 313

\bibitem[{{Chu} \& {Mac Low}(1990)}]{Chu_MacLow_1990}
{Chu}, Y.-H., \& {Mac Low}, M.-M. 1990,
  \href{http://dx.doi.org/10.1086/169505}{\JournalTitle{\apj}, 365, 510}

\bibitem[{{Cohen} {et~al.}(2003){Cohen}, {Staveley-Smith}, \&
  {Green}}]{Cohen_etal_2003}
{Cohen}, M., {Staveley-Smith}, L., \& {Green}, A. 2003,
  \href{http://dx.doi.org/10.1046/j.1365-8711.2003.06303.x}{\JournalTitle{\mnras}, 340, 275}

\bibitem[{{Davies} {et~al.}(1976){Davies}, {Elliott}, \&
  {Meaburn}}]{Davies_Meaburn_1976}
{Davies}, R.~D., {Elliott}, K.~H., \& {Meaburn}, J. 1976,
  \JournalTitle{\memras}, 81, 89

\bibitem[{{Dawson}(2013)}]{Dawson_2013}
{Dawson}, J.~R. 2013,
  \href{http://dx.doi.org/10.1017/pas.2013.002}{\JournalTitle{\pasa}, 30, 25}

\bibitem[{{Dawson} {et~al.}(2011){Dawson}, {McClure-Griffiths}, {Kawamura},
  {Mizuno}, {Onishi}, {Mizuno}, \& {Fukui}}]{Dawson_etal_2011b}
{Dawson}, J.~R., {McClure-Griffiths}, N.~M., {Kawamura}, A., {et~al.} 2011,
  \href{http://dx.doi.org/10.1088/0004-637X/728/2/127}{\JournalTitle{\apj},
  728, 127}

\bibitem[{{Dawson} {et~al.}(2013){Dawson}, {McClure-Griffiths}, {Wong},
  {Dickey}, {Hughes}, {Fukui}, \& {Kawamura}}]{Dawson_etal_2013}
{Dawson}, J.~R., {McClure-Griffiths}, N.~M., {Wong}, T., {et~al.} 2013,
  \href{http://dx.doi.org/10.1088/0004-637X/763/1/56}{\JournalTitle{\apj}, 763,
  56}

\bibitem[{{Dickey} {et~al.}(1994){Dickey}, {Mebold}, {Marx}, {Amy}, {Haynes},
  \& {Wilson}}]{Dickey_etal_1994}
{Dickey}, J.~M., {Mebold}, U., {Marx}, M., {et~al.} 1994, \JournalTitle{\aap},
  289, 357

\bibitem[{{Dobashi} {et~al.}(1996){Dobashi}, {Bernard}, \&
  {Fukui}}]{Dobashi_etal_1996}
{Dobashi}, K., {Bernard}, J.-P., \& {Fukui}, Y. 1996,
  \href{http://dx.doi.org/10.1086/177509}{\JournalTitle{\apj}, 466, 282}

\bibitem[{{Domgoergen} {et~al.}(1995){Domgoergen}, {Bomans}, \& {de
  Boer}}]{Domgorgen_etal_1995}
{Domgoergen}, H., {Bomans}, D.~J., \& {de Boer}, K.~S. 1995,
  \JournalTitle{\aap}, 296, 523

\bibitem[{{Dopita} {et~al.}(1985){Dopita}, {Mathewson}, \&
  {Ford}}]{Dopita_etal_1985}
{Dopita}, M.~A., {Mathewson}, D.~S., \& {Ford}, V.~L. 1985,
  \href{http://dx.doi.org/10.1086/163556}{\JournalTitle{\apj}, 297, 599}

\bibitem[{{Efremov} \& {Elmegreen}(1998)}]{Efremov_Elmegreen_1998}
{Efremov}, Y.~N., \& {Elmegreen}, B.~G. 1998,
  \href{http://dx.doi.org/10.1046/j.1365-8711.1998.01745.x}{\JournalTitle{\mnras}, 299, 643}

\bibitem[{{Efremov} {et~al.}(1998){Efremov}, {Elmegreen}, \&
  {Hodge}}]{Efremov_etal_1998}
{Efremov}, Y.~N., {Elmegreen}, B.~G., \& {Hodge}, P.~W. 1998,
  \href{http://dx.doi.org/10.1086/311468}{\JournalTitle{\apjl}, 501, L163}

\bibitem[{{Ehlerova} {et~al.}(1997){Ehlerova}, {Palous}, {Theis}, \&
  {Hensler}}]{Ehlerova_etal_1997}
{Ehlerova}, S., {Palous}, J., {Theis}, C., \& {Hensler}, G. 1997,
  \JournalTitle{\aap}, 328, 121

\bibitem[{{Elmegreen}(1998)}]{Elmegreen_1998}
{Elmegreen}, B.~G. 1998, in Astronomical Society of the Pacific Conference
  Series, Vol. 148, Origins, ed. C.~E. {Woodward}, J.~M. {Shull}, \& H.~A.
  {Thronson}, Jr., 150

\bibitem[{{Emerson} \& {Graeve}(1988)}]{Emerson_Graeve_1988}
{Emerson}, D.~T., \& {Graeve}, R. 1988, \JournalTitle{\aap}, 190, 353

\bibitem[{{Ezawa} {et~al.}(2004){Ezawa}, {Kawabe}, {Kohno}, \&
  {Yamamoto}}]{Ezawa_etal_2004}
{Ezawa}, H., {Kawabe}, R., {Kohno}, K., \& {Yamamoto}, S. 2004,
  \href{http://dx.doi.org/10.1117/12.551391}{in Society of Photo-Optical
  Instrumentation Engineers (SPIE) Conference Series, Vol. 5489, Society of
  Photo-Optical Instrumentation Engineers (SPIE) Conference Series, ed. J.~M.
  {Oschmann}, Jr.}, 763

\bibitem[{{Ezawa} {et~al.}(2008){Ezawa}, {Kohno}, {Kawabe}, {Yamamoto},
  {Inoue}, {Iwashita}, {Matsuo}, {Okuda}, {Oshima}, {Sakai}, {Tanaka},
  {Yamaguchi}, {Wilson}, {Yun}, {Aretxaga}, {Hughes}, {Austermann}, {Perera},
  {Scott}, {Bronfman}, \& {Cortes}}]{Ezawa_etal_2008}
{Ezawa}, H., {Kohno}, K., {Kawabe}, R., {et~al.} 2008,
  \href{http://dx.doi.org/10.1117/12.789652}{in Society of Photo-Optical
  Instrumentation Engineers (SPIE) Conference Series, Vol. 7012, Society of
  Photo-Optical Instrumentation Engineers (SPIE) Conference Series}

\bibitem[{{Fazio} {et~al.}(2004){Fazio}, {Hora}, {Allen}, {Ashby}, {Barmby},
  {Deutsch}, {Huang}, {Kleiner}, {Marengo}, {Megeath}, {Melnick}, {Pahre},
  {Patten}, {Polizotti}, {Smith}, {Taylor}, {Wang}, {Willner}, {Hoffmann},
  {Pipher}, {Forrest}, {McMurty}, {McCreight}, {McKelvey}, {McMurray}, {Koch},
  {Moseley}, {Arendt}, {Mentzell}, {Marx}, {Losch}, {Mayman}, {Eichhorn},
  {Krebs}, {Jhabvala}, {Gezari}, {Fixsen}, {Flores}, {Shakoorzadeh}, {Jungo},
  {Hakun}, {Workman}, {Karpati}, {Kichak}, {Whitley}, {Mann}, {Tollestrup},
  {Eisenhardt}, {Stern}, {Gorjian}, {Bhattacharya}, {Carey}, {Nelson},
  {Glaccum}, {Lacy}, {Lowrance}, {Laine}, {Reach}, {Stauffer}, {Surace},
  {Wilson}, {Wright}, {Hoffman}, {Domingo}, \& {Cohen}}]{Fazio_etal_2004}
{Fazio}, G.~G., {Hora}, J.~L., {Allen}, L.~E., {et~al.} 2004,
  \href{http://dx.doi.org/10.1086/422843}{\JournalTitle{\apjs}, 154, 10}

\bibitem[{{Filipovic} {et~al.}(1995){Filipovic}, {Haynes}, {White}, {Jones},
  {Klein}, \& {Wielebinski}}]{Filipovic_etal_1995}
{Filipovic}, M.~D., {Haynes}, R.~F., {White}, G.~L., {et~al.} 1995,
  \JournalTitle{\aaps}, 111, 311

\bibitem[{{Filipovi{\'c}} \&
  {Staveley-Smith}(1998)}]{Filipovic_Staveley-Smith_1998}
{Filipovi{\'c}}, M.~D., \& {Staveley-Smith}, L. 1998, in The Magellanic Clouds
  and Other Dwarf Galaxies, ed. T.~{Richtler} \& J.~M. {Braun}, 137

\bibitem[{{Fukui} \& {Kawamura}(2010)}]{Fukui_Kawamura_2010}
{Fukui}, Y., \& {Kawamura}, A. 2010,
  \href{http://dx.doi.org/10.1146/annurev-astro-081309-130854}{\JournalTitle{\%
araa}, 48, 547}

\bibitem[{{Fukui} {et~al.}(2008){Fukui}, {Kawamura}, {Minamidani}, {Mizuno},
  {Kanai}, {Mizuno}, {Onishi}, {Yonekura}, {Mizuno}, {Ogawa}, \&
  {Rubio}}]{Fukui_etal_2008}
{Fukui}, Y., {Kawamura}, A., {Minamidani}, T., {et~al.} 2008,
  \href{http://dx.doi.org/10.1086/589833}{\JournalTitle{\apjs}, 178, 56}

\bibitem[{{Fukui} {et~al.}(2009){Fukui}, {Kawamura}, {Wong}, {Murai},
  {Iritani}, {Mizuno}, {Mizuno}, {Onishi}, {Hughes}, {Ott}, {Muller},
  {Staveley-Smith}, \& {Kim}}]{Fukui_etal_2009}
{Fukui}, Y., {Kawamura}, A., {Wong}, T., {et~al.} 2009,
  \href{http://dx.doi.org/10.1088/0004-637X/705/1/144}{\JournalTitle{\apj},
  705, 144}

\bibitem[{{Glatt} {et~al.}(2010){Glatt}, {Grebel}, \& {Koch}}]{Glatt_etal_2010}
{Glatt}, K., {Grebel}, E.~K., \& {Koch}, A. 2010,
  \href{http://dx.doi.org/10.1051/0004-6361/201014187}{\JournalTitle{\aap},
  517, A50}

\bibitem[{{Goldreich} \& {Kwan}(1974)}]{Goldreich_Kwan_1974}
{Goldreich}, P., \& {Kwan}, J. 1974,
  \href{http://dx.doi.org/10.1086/152821}{\JournalTitle{\apj}, 189, 441}

\bibitem[{{Gruendl} \& {Chu}(2009)}]{Gruendl_Chu_2009}
{Gruendl}, R.~A., \& {Chu}, Y.-H. 2009,
  \href{http://dx.doi.org/10.1088/0067-0049/184/1/172}{\JournalTitle{\apjs},
  184, 172}

\bibitem[{{Hartmann} {et~al.}(2001){Hartmann}, {Ballesteros-Paredes}, \&
  {Bergin}}]{Hartmann_etal_2001}
{Hartmann}, L., {Ballesteros-Paredes}, J., \& {Bergin}, E.~A. 2001,
  \href{http://dx.doi.org/10.1086/323863}{\JournalTitle{\apj}, 562, 852}

\bibitem[{{Henize}(1956)}]{Henize_1956}
{Henize}, K.~G. 1956,
  \href{http://dx.doi.org/10.1086/190025}{\JournalTitle{\apjs}, 2, 315}

\bibitem[{{Hosokawa} \& {Inutsuka}(2007)}]{Hosokawa_Inutsuka_2007}
{Hosokawa}, T., \& {Inutsuka}, S.-i. 2007,
  \href{http://dx.doi.org/10.1086/518396}{\JournalTitle{\apj}, 664, 363}

\bibitem[{{Hughes} {et~al.}(2007){Hughes}, {Staveley-Smith}, {Kim}, {Wolleben},
  \& {Filipovi{\'c}}}]{Hughes_etal_2007}
{Hughes}, A., {Staveley-Smith}, L., {Kim}, S., {Wolleben}, M., \&
  {Filipovi{\'c}}, M. 2007,
  \href{http://dx.doi.org/10.1111/j.1365-2966.2007.12466.x}{\JournalTitle{\mnras}, 382, 543}

\bibitem[{{Ikeda} {et~al.}(2009){Ikeda}, {Kitamura}, \&
  {Sunada}}]{Ikeda_etal_2009}
{Ikeda}, N., {Kitamura}, Y., \& {Sunada}, K. 2009,
  \href{http://dx.doi.org/10.1088/0004-637X/691/2/1560}{\JournalTitle{\apj},
  691, 1560}

\bibitem[{{Ikeda} {et~al.}(2007){Ikeda}, {Sunada}, \&
  {Kitamura}}]{Ikeda_etal_2007}
{Ikeda}, N., {Sunada}, K., \& {Kitamura}, Y. 2007,
  \href{http://dx.doi.org/10.1086/519484}{\JournalTitle{\apj}, 665, 1194}

\bibitem[{{Indebetouw} {et~al.}(2013){Indebetouw}, {Brogan}, {Chen}, {Leroy},
  {Johnson}, {Muller}, {Madden}, {Cormier}, {Galliano}, {Hughes}, {Hunter},
  {Kawamura}, {Kepley}, {Lebouteiller}, {Meixner}, {Oliveira}, {Onishi}, \&
  {Vasyunina}}]{Indebetouw_etal_2013}
{Indebetouw}, R., {Brogan}, C., {Chen}, C.-H.~R., {et~al.} 2013,
  \href{http://dx.doi.org/10.1088/0004-637X/774/1/73}{\JournalTitle{\apj}, 774,
  73}

\bibitem[{{Inoue} {et~al.}(2008){Inoue}, {Muraoka}, {Sakai}, {Endo}, {Kohno},
  {Asayama}, {Noguchi}, \& {Ogawa}}]{Inoue_etal_2008}
{Inoue}, H., {Muraoka}, K., {Sakai}, T., {et~al.} 2008, in Ninteenth
  International Symposium on Space Terahertz Technology, ed. W.~{Wild}, 281

\bibitem[{{Jones} {et~al.}(1996){Jones}, {Tielens}, \&
  {Hollenbach}}]{Jones_etal_1996}
{Jones}, A.~P., {Tielens}, A.~G.~G.~M., \& {Hollenbach}, D.~J. 1996,
  \href{http://dx.doi.org/10.1086/177823}{\JournalTitle{\apj}, 469, 740}

\bibitem[{{Kawamura} {et~al.}(1998){Kawamura}, {Onishi}, {Yonekura}, {Dobashi},
  {Mizuno}, {Ogawa}, \& {Fukui}}]{Kawamura_etal_1998}
{Kawamura}, A., {Onishi}, T., {Yonekura}, Y., {et~al.} 1998,
  \href{http://dx.doi.org/10.1086/313119}{\JournalTitle{\apjs}, 117, 387}

\bibitem[{{Kawamura} {et~al.}(2009){Kawamura}, {Mizuno}, {Minamidani},
  {Filipovi{\'c}}, {Staveley-Smith}, {Kim}, {Mizuno}, {Onishi}, {Mizuno}, \&
  {Fukui}}]{Kawamura_etal_2009}
{Kawamura}, A., {Mizuno}, Y., {Minamidani}, T., {et~al.} 2009,
  \href{http://dx.doi.org/10.1088/0067-0049/184/1/1}{\JournalTitle{\apjs}, 184,
  1}

\bibitem[{{Kim} {et~al.}(1999){Kim}, {Dopita}, {Staveley-Smith}, \&
  {Bessell}}]{Kim_etal_1999}
{Kim}, S., {Dopita}, M.~A., {Staveley-Smith}, L., \& {Bessell}, M.~S. 1999,
  \href{http://dx.doi.org/10.1086/301116}{\JournalTitle{\aj}, 118, 2797}

\bibitem[{{Kim} {et~al.}(1998){Kim}, {Staveley-Smith}, {Dopita}, {Freeman},
  {Sault}, {Kesteven}, \& {McConnell}}]{Kim_etal_1998}
{Kim}, S., {Staveley-Smith}, L., {Dopita}, M.~A., {et~al.} 1998,
  \href{http://dx.doi.org/10.1086/306030}{\JournalTitle{\apj}, 503, 674}

\bibitem[{{Kim} {et~al.}(2003){Kim}, {Staveley-Smith}, {Dopita}, {Sault},
  {Freeman}, {Lee}, \& {Chu}}]{Kim_etal_2003}
---. 2003, \href{http://dx.doi.org/10.1086/376980}{\JournalTitle{\apjs}, 148,
  473}

\bibitem[{{Kohno}(2005)}]{Kohno_2005}
{Kohno}, K. 2005, in Astronomical Society of the Pacific Conference Series,
  Vol. 344, The Cool Universe: Observing Cosmic Dawn, ed. C.~{Lidman} \&
  D.~{Alloin}, 242

\bibitem[{{Koyama} \& {Inutsuka}(2000)}]{Koyama_Inutsuka_2000}
{Koyama}, H., \& {Inutsuka}, S.-I. 2000,
  \href{http://dx.doi.org/10.1086/308594}{\JournalTitle{\apj}, 532, 980}

\bibitem[{{Ladd} {et~al.}(2005){Ladd}, {Purcell}, {Wong}, \&
  {Robertson}}]{Ladd_etal_2005}
{Ladd}, N., {Purcell}, C., {Wong}, T., \& {Robertson}, S. 2005,
  \href{http://dx.doi.org/10.1071/AS04068}{\JournalTitle{\pasa}, 22, 62}

\bibitem[{{Li} \& {Draine}(2001)}]{Li_Draine_2001}
{Li}, A., \& {Draine}, B.~T. 2001,
  \href{http://dx.doi.org/10.1086/323147}{\JournalTitle{\apj}, 554, 778}

\bibitem[{{Mac Low} \& {Ferrara}(1999)}]{MacLow_Ferrara_1999}
{Mac Low}, M.-M., \& {Ferrara}, A. 1999,
  \href{http://dx.doi.org/10.1086/306832}{\JournalTitle{\apj}, 513, 142}

\bibitem[{{Mac Low} {et~al.}(1989){Mac Low}, {McCray}, \&
  {Norman}}]{MacLow_etal_1989}
{Mac Low}, M.-M., {McCray}, R., \& {Norman}, M.~L. 1989,
  \href{http://dx.doi.org/10.1086/167094}{\JournalTitle{\apj}, 337, 141}

\bibitem[{{MacLaren} {et~al.}(1988){MacLaren}, {Richardson}, \&
  {Wolfendale}}]{MacLaren_etal_1988}
{MacLaren}, I., {Richardson}, K.~M., \& {Wolfendale}, A.~W. 1988,
  \href{http://dx.doi.org/10.1086/166791}{\JournalTitle{\apj}, 333, 821}

\bibitem[{{Marx-Zimmer} {et~al.}(2000){Marx-Zimmer}, {Herbstmeier}, {Dickey},
  {Zimmer}, {Staveley-Smith}, \& {Mebold}}]{MarxZimmer_etal_2000}
{Marx-Zimmer}, M., {Herbstmeier}, U., {Dickey}, J.~M., {et~al.} 2000,
  \JournalTitle{\aap}, 354, 787

\bibitem[{{McCray} \& {Kafatos}(1987)}]{McCray_Kafatos_1987}
{McCray}, R., \& {Kafatos}, M. 1987,
  \href{http://dx.doi.org/10.1086/165267}{\JournalTitle{\apj}, 317, 190}

\bibitem[{{Meaburn}(1980)}]{Meaburn_1980}
{Meaburn}, J. 1980, \JournalTitle{\mnras}, 192, 365

\bibitem[{{Meixner} {et~al.}(2006){Meixner}, {Gordon}, {Indebetouw}, {Hora},
  {Whitney}, {Blum}, {Reach}, {Bernard}, {Meade}, {Babler}, {Engelbracht},
  {For}, {Misselt}, {Vijh}, {Leitherer}, {Cohen}, {Churchwell}, {Boulanger},
  {Frogel}, {Fukui}, {Gallagher}, {Gorjian}, {Harris}, {Kelly}, {Kawamura},
  {Kim}, {Latter}, {Madden}, {Markwick-Kemper}, {Mizuno}, {Mizuno}, {Mould},
  {Nota}, {Oey}, {Olsen}, {Onishi}, {Paladini}, {Panagia}, {Perez-Gonzalez},
  {Shibai}, {Sato}, {Smith}, {Staveley-Smith}, {Tielens}, {Ueta}, {van Dyk},
  {Volk}, {Werner}, \& {Zaritsky}}]{Meixner_etal_2006}
{Meixner}, M., {Gordon}, K.~D., {Indebetouw}, R., {et~al.} 2006,
  \href{http://dx.doi.org/10.1086/508185}{\JournalTitle{\aj}, 132, 2268}

\bibitem[{{Minamidani} {et~al.}(2008){Minamidani}, {Mizuno}, Y., {Kawamura},
  {Onishi}, {Hasegawa}, {Tatematsu}, {Ikeda}, {Moriguchi}, {Yamaguchi}, {Ott},
  {Wong}, {Muller}, {Pineda}, {Hughes}, {Staveley-Smith}, {Klein}, {Mizuno},
  {Nikoli{\'c}}, {Booth}, {Heikkil{\"a}}, {Nyman}, {Lerner}, {Garay}, {Kim},
  {Fujishita}, {Kawase}, {Rubio}, \& {Fukui}}]{Minamidani_etal_2008}
{Minamidani}, T., {Mizuno}, N., Y., M., {et~al.} 2008,
  \href{http://dx.doi.org/10.1086/524038}{\JournalTitle{\apjs}, 175, 485}

\bibitem[{{Minamidani} {et~al.}(2011){Minamidani}, {Tanaka}, {Mizuno},
  {Mizuno}, {Kawamura}, {Onishi}, {Hasegawa}, {Tatematsu}, {Takekoshi},
  {Sorai}, {Moribe}, {Torii}, {Sakai}, {Muraoka}, {Tanaka}, {Ezawa}, {Kohno},
  {Kim}, {Rubio}, \& {Fukui}}]{Minamidani_etal_2011}
{Minamidani}, T., {Tanaka}, T., {Mizuno}, Y., {et~al.} 2011,
  \href{http://dx.doi.org/10.1088/0004-6256/141/3/73}{\JournalTitle{\aj}, 141,
  73}

\bibitem[{{Mizuno} {et~al.}(2001){Mizuno}, {Yamaguchi}, {Mizuno}, {Rubio},
  {Abe}, {Saito}, {Onishi}, {Yonekura}, {Yamaguchi}, {Ogawa}, \&
  {Fukui}}]{Mizuno_etal_2001}
{Mizuno}, N., {Yamaguchi}, R., {Mizuno}, A., {et~al.} 2001,
  \href{http://dx.doi.org/10.1093/pasj/53.6.971}{\JournalTitle{\pasj}, 53, 971}

\bibitem[{{Mizuno} {et~al.}(2010){Mizuno}, {Kawamura}, {Onishi}, {Minamidani},
  {Muller}, {Yamamoto}, {Hayakawa}, {Mizuno}, {Mizuno}, {Stutzki}, {Pineda},
  {Klein}, {Bertoldi}, {Koo}, {Rubio}, {Burton}, {Benz}, {Ezawa}, {Yamaguchi},
  {Kohno}, {Hasegawa}, {Tatematsu}, {Ikeda}, {Ott}, {Wong}, {Hughes},
  {Meixner}, {Indebetouw}, {Gordon}, {Whitney}, {Bernard}, \&
  {Fukui}}]{Mizuno_etal_2010}
{Mizuno}, Y., {Kawamura}, A., {Onishi}, T., {et~al.} 2010,
  \JournalTitle{\pasj}, 62, 51

\bibitem[{{Ntormousi} {et~al.}(2011){Ntormousi}, {Burkert}, {Fierlinger}, \&
  {Heitsch}}]{Ntormousi_etal_2011}
{Ntormousi}, E., {Burkert}, A., {Fierlinger}, K., \& {Heitsch}, F. 2011,
  \href{http://dx.doi.org/10.1088/0004-637X/731/1/13}{\JournalTitle{\apj}, 731,
  13}

\bibitem[{{P{\'e}rez-Gonz{\'a}lez} {et~al.}(2006){P{\'e}rez-Gonz{\'a}lez},
  {Kennicutt}, {Gordon}, {Misselt}, {Gil de Paz}, {Engelbracht}, {Rieke},
  {Bendo}, {Bianchi}, {Boissier}, {Calzetti}, {Dale}, {Draine}, {Jarrett},
  {Hollenbach}, \& {Prescott}}]{Perez-Gonzalez_etal_2006}
{P{\'e}rez-Gonz{\'a}lez}, P.~G., {Kennicutt}, Jr., R.~C., {Gordon}, K.~D.,
  {et~al.} 2006, \href{http://dx.doi.org/10.1086/506196}{\JournalTitle{\apj},
  648, 987}

\bibitem[{{Pietrzy{\'n}ski} {et~al.}(2013){Pietrzy{\'n}ski}, {Graczyk},
  {Gieren}, {Thompson}, {Pilecki}, {Udalski}, {Soszy{\'n}ski}, {Koz{\l}owski},
  {Konorski}, {Suchomska}, {Bono}, {Moroni}, {Villanova}, {Nardetto},
  {Bresolin}, {Kudritzki}, {Storm}, {Gallenne}, {Smolec}, {Minniti}, {Kubiak},
  {Szyma{\'n}ski}, {Poleski}, {Wyrzykowski}, {Ulaczyk}, {Pietrukowicz},
  {G{\'o}rski}, \& {Karczmarek}}]{Pietrynski_etal_2013}
{Pietrzy{\'n}ski}, G., {Graczyk}, D., {Gieren}, W., {et~al.} 2013,
  \href{http://dx.doi.org/10.1038/nature11878}{\JournalTitle{\nat}, 495, 76}

\bibitem[{{Points} {et~al.}(1999){Points}, {Chu}, {Kim}, {Smith}, {Snowden},
  {Brandner}, \& {Gruendl}}]{Points_etal_1999}
{Points}, S.~D., {Chu}, Y.~H., {Kim}, S., {et~al.} 1999,
  \href{http://dx.doi.org/10.1086/307249}{\JournalTitle{\apj}, 518, 298}

\bibitem[{{Rieke} {et~al.}(2004){Rieke}, {Young}, {Cadien}, {Engelbracht},
  {Gordon}, {Kelly}, {Low}, {Misselt}, {Morrison}, {Muzerolle}, {Rivlis},
  {Stansberry}, {Beeman}, {Haller}, {Frayer}, {Latter}, {Noriega-Crespo},
  {Padgett}, {Hines}, {Bean}, {Burmester}, {Heim}, {Glenn}, {Ordonez},
  {Schwenker}, {Siewert}, {Strecker}, {Tennant}, {Troeltzsch}, {Unruh},
  {Warden}, {Ade}, {Alonso-Herrero}, {Blaylock}, {Dole}, {Egami}, {Hinz}, {Le
  Floc'h}, {Papovich}, {Perez-Gonzalez}, {Rieke}, {Smith}, {Su}, {Bennett},
  {Henderson}, {Lu}, {Masci}, {Pesenson}, {Rebull}, {Rho}, {Keene}, {Stolovy},
  {Wachter}, {Wheaton}, {Richards}, {Garner}, {Hegge}, {Henderson}, {MacFeely},
  {Michika}, {Miller}, {Neitenbach}, {Winghart}, {Woodruff}, {Arens},
  {Beichman}, {Gaalema}, {Gautier}, {Lada}, {Mould}, {Neugebauer}, \&
  {Stapelfeldt}}]{Rieke_etal_2004}
{Rieke}, G.~H., {Young}, E.~T., {Cadien}, J., {et~al.} 2004,
  \href{http://dx.doi.org/10.1117/12.551965}{in Society of Photo-Optical
  Instrumentation Engineers (SPIE) Conference Series, Vol. 5487, Society of
  Photo-Optical Instrumentation Engineers (SPIE) Conference Series, ed. J.~C.
  {Mather}}, 50

\bibitem[{{Sawada} {et~al.}(2008){Sawada}, {Ikeda}, {Sunada}, {Kuno},
  {Kamazaki}, {Morita}, {Kurono}, {Koura}, {Abe}, {Kawase}, {Maekawa},
  {Horigome}, \& {Yanagisawa}}]{Sawada_etal_2008}
{Sawada}, T., {Ikeda}, N., {Sunada}, K., {et~al.} 2008,
  \href{http://dx.doi.org/10.1093/pasj/60.3.445}{\JournalTitle{\pasj}, 60, 445}

\bibitem[{{Sch{\"o}ier} {et~al.}(2005){Sch{\"o}ier}, {van der Tak}, {van
  Dishoeck}, \& {Black}}]{Schoier_etal_2005}
{Sch{\"o}ier}, F.~L., {van der Tak}, F.~F.~S., {van Dishoeck}, E.~F., \&
  {Black}, J.~H. 2005,
  \href{http://dx.doi.org/10.1051/0004-6361:20041729}{\JournalTitle{\aap}, 432,
  369}

\bibitem[{{Scoville} \& {Solomon}(1974)}]{Scoville_Solomon_1974}
{Scoville}, N.~Z., \& {Solomon}, P.~M. 1974,
  \href{http://dx.doi.org/10.1086/181398}{\JournalTitle{\apjl}, 187, L67}

\bibitem[{{Seale} {et~al.}(2012){Seale}, {Looney}, {Wong}, {Ott}, {Klein}, \&
  {Pineda}}]{Seale_etal_2012}
{Seale}, J.~P., {Looney}, L.~W., {Wong}, T., {et~al.} 2012,
  \href{http://dx.doi.org/10.1088/0004-637X/751/1/42}{\JournalTitle{\apj}, 751,
  42}

\bibitem[{{Smith} \& {MCELS Team}(1999)}]{MCELS_1999}
{Smith}, R.~C., \& {MCELS Team}. 1999, in IAU Symposium, Vol. 190, New Views of
  the Magellanic Clouds, ed. Y.-H. {Chu}, N.~{Suntzeff}, J.~{Hesser}, \&
  D.~{Bohlender}, 28

\bibitem[{{Sorai} {et~al.}(2000){Sorai}, {Sunada}, {Okumura}, {Tetsuro},
  {Tanaka}, {Natori}, \& {Onuki}}]{Sorai_etal_2000}
{Sorai}, K., {Sunada}, K., {Okumura}, S.~K., {et~al.} 2000, in Society of
  Photo-Optical Instrumentation Engineers (SPIE) Conference Series, Vol. 4015,
  Society of Photo-Optical Instrumentation Engineers (SPIE) Conference Series,
  ed. H.~R. {Butcher}, 86

\bibitem[{{Staveley-Smith} {et~al.}(2003){Staveley-Smith}, {Kim}, {Calabretta},
  {Haynes}, \& {Kesteven}}]{Staveley-Smith_etal_2003}
{Staveley-Smith}, L., {Kim}, S., {Calabretta}, M.~R., {Haynes}, R.~F., \&
  {Kesteven}, M.~J. 2003,
  \href{http://dx.doi.org/10.1046/j.1365-8711.2003.06146.x}{\JournalTitle{\mnras}, 339, 87}

\bibitem[{{Tenorio-Tagle} \& {Bodenheimer}(1988)}]{Tenorio-Tagle_etal_1988}
{Tenorio-Tagle}, G., \& {Bodenheimer}, P. 1988,
  \href{http://dx.doi.org/10.1146/annurev.aa.26.090188.001045}{\JournalTitle{\%
araa}, 26, 145}

\bibitem[{{van der Marel} \& {Cioni}(2001)}]{vanderMarel_Cioni_2001}
{van der Marel}, R.~P., \& {Cioni}, M.-R.~L. 2001,
  \href{http://dx.doi.org/10.1086/323099}{\JournalTitle{\aj}, 122, 1807}

\bibitem[{{Walter} \& {Brinks}(1999{\natexlab{a}})}]{Walter_etal_1999}
{Walter}, F., \& {Brinks}, E. 1999{\natexlab{a}},
  \href{http://dx.doi.org/10.1086/300906}{\JournalTitle{\aj}, 118, 273}

\bibitem[{{Walter} \& {Brinks}(1999{\natexlab{b}})}]{Walter_Brinks_1999}
---. 1999{\natexlab{b}},
  \href{http://dx.doi.org/10.1086/300906}{\JournalTitle{\aj}, 118, 273}

\bibitem[{{Weisz} {et~al.}(2009){Weisz}, {Skillman}, {Cannon}, {Walter},
  {Brinks}, {Ott}, \& {Dolphin}}]{Weisz_etal_2009}
{Weisz}, D.~R., {Skillman}, E.~D., {Cannon}, J.~M., {et~al.} 2009,
  \href{http://dx.doi.org/10.1088/0004-637X/691/1/L59}{\JournalTitle{\apjl},
  691, L59}

\bibitem[{{Whitney} {et~al.}(2008){Whitney}, {Sewilo}, {Indebetouw},
  {Robitaille}, {Meixner}, {Gordon}, {Meade}, {Babler}, {Harris}, {Hora},
  {Bracker}, {Povich}, {Churchwell}, {Engelbracht}, {For}, {Block}, {Misselt},
  {Vijh}, {Leitherer}, {Kawamura}, {Blum}, {Cohen}, {Fukui}, {Mizuno},
  {Mizuno}, {Srinivasan}, {Tielens}, {Volk}, {Bernard}, {Boulanger}, {Frogel},
  {Gallagher}, {Gorjian}, {Kelly}, {Latter}, {Madden}, {Kemper}, {Mould},
  {Nota}, {Oey}, {Olsen}, {Onishi}, {Paladini}, {Panagia}, {Perez-Gonzalez},
  {Reach}, {Shibai}, {Sato}, {Smith}, {Staveley-Smith}, {Ueta}, {Van Dyk},
  {Werner}, {Wolff}, \& {Zaritsky}}]{Whitney_etal_2008}
{Whitney}, B.~A., {Sewilo}, M., {Indebetouw}, R., {et~al.} 2008,
  \href{http://dx.doi.org/10.1088/0004-6256/136/1/18}{\JournalTitle{\aj}, 136,
  18}

\bibitem[{{Will} {et~al.}(1996){Will}, {Bomans}, {Vallenari}, {Schmidt}, \& {de
  Boer}}]{Will_etal_1996}
{Will}, J.-M., {Bomans}, D.~J., {Vallenari}, A., {Schmidt}, J.~H.~K., \& {de
  Boer}, K.~S. 1996, \JournalTitle{\aap}, 315, 125

\bibitem[{{Wong} {et~al.}(2011){Wong}, {Hughes}, {Ott}, {Muller}, {Pineda},
  {Bernard}, {Chu}, {Fukui}, {Gruendl}, {Henkel}, {Kawamura}, {Klein},
  {Looney}, {Maddison}, {Mizuno}, {Paradis}, {Seale}, \&
  {Welty}}]{Wong_etal_2011}
{Wong}, T., {Hughes}, A., {Ott}, J., {et~al.} 2011,
  \href{http://dx.doi.org/10.1088/0067-0049/197/2/16}{\JournalTitle{\apjs},
  197, 16}

\bibitem[{{Yamaguchi} {et~al.}(2001{\natexlab{a}}){Yamaguchi}, {Mizuno},
  {Onishi}, {Mizuno}, \& {Fukui}}]{Yamaguchi_etal_2001a}
{Yamaguchi}, R., {Mizuno}, N., {Onishi}, T., {Mizuno}, A., \& {Fukui}, Y.
  2001{\natexlab{a}},
  \href{http://dx.doi.org/10.1086/320678}{\JournalTitle{\apjl}, 553, L185}

\bibitem[{{Yamaguchi} {et~al.}(2001{\natexlab{b}}){Yamaguchi}, {Mizuno},
  {Onishi}, {Mizuno}, \& {Fukui}}]{Yamaguchi_etal_2001b}
---. 2001{\natexlab{b}}, \JournalTitle{\pasj}, 53, 959

\bibitem[{{Yonekura} {et~al.}(1997){Yonekura}, {Dobashi}, {Mizuno}, {Ogawa}, \&
  {Fukui}}]{Yonekura_etal_1997}
{Yonekura}, Y., {Dobashi}, K., {Mizuno}, A., {Ogawa}, H., \& {Fukui}, Y. 1997,
  \href{http://dx.doi.org/10.1086/312994}{\JournalTitle{\apjs}, 110, 21}

\end{thebibliography}

\clearpage
\begin{deluxetable}{lccc}
\tablecolumns{4}
\tablewidth{0pc}
\tabletypesize{\scriptsize}
\tablecaption{Observation parameters}
\tablehead{
& \multicolumn{3}{c}{Values}\\
\cline{2-4}\\
\multicolumn{1}{c}{Parameters} & \multicolumn{2}{c}{$^{12}$CO($J=$ 3--2)} & $^{12}$CO($J=$ 1--0),$^{13}$CO($J=$ 1--0) }
\startdata
Observation Period & 2006 Nov. & 2011 Nov. & 2012 Jul. \\
Telescope & ASTE & ASTE & Mopra\\
$\;\;\;\;\;\;\;\;$Diameter & 10 m  & 10 m  & 22 m \\
$\;\;\;\;\;\;\;\;$Rest frequency　& 345.79599 GHz & 345.79599 GHz & 115.27120 GHz ($^{12}$CO) \\
&&& 110.20135 GHz ($^{13}$CO)\\
$\;\;\;\;\;\;\;\;$Beam size(HPBW) & 22$^{\prime \prime}$ & 22$^{\prime \prime}$ & 33$^{\prime \prime}$\\
Spectrometer & MAC & MAC & MOPS\\
$\;\;\;\;\;\;\;\;$Channel number & 1024 & 1024 & 4096\\
$\;\;\;\;\;\;\;\;$Bandwidth & 512 MHz (440 km s$^{-1}$) & 128 MHz (125 km s$^{-1}$) & 138 MHz (376 km s$^{-1}$)\\
$\;\;\;\;\;\;\;\;$Channel spacing & 0.5 MHz (0.44 km s$^{-1}$) & 0.125 MHz (0.1 km s$^{-1}$) & 0.03 MHz (0.09 km s$^{-1}$)\\
Map description & & & \\
$\;\;\;\;\;\;\;\;$Mapping mode & On-the-fly & On-the-fly & On-the-fly \\
$\;\;\;\;\;\;\;\;$Target region & N49   & N48  & N48,N49/Each clump\\ 
$\;\;\;\;\;\;\;\;$Field coverage & 3$^{\prime} \times$ 5.5$^{\prime}$ & 
   		7$^{\prime } \times$ 7$^{\prime } \times$ 2 fields  &
		2$^{\prime } \times$ 2$^{\prime } \times$ 11 fields \\
$\;\;\;\;\;\;\;\;$Effective beam width & 27$^{\prime \prime}$ & 27$^{\prime \prime}$ & 45$^{\prime \prime}$\\
$\;\;\;\;\;\;\;\;$rms noise level ($\Delta v = 0.44$ km s$^{-1}$) & 0.4 K & 0.3 K &  0.1\phantom{0} K($^{12}$CO) \\
   &&& 0.04 K($^{13}$CO)
\enddata
\label{tab_obs}
\end{deluxetable}


\begin{deluxetable}{ccccccccc}[ht]
\tablecolumns{9}
\tablewidth{0pc}
\tabletypesize{\scriptsize}
\tablecaption{Parameters of $^{12}$CO($J=$3--2) integrated intensity local peaks}
\tablehead{
&\multicolumn{2}{c}{Peak Position}& &\multicolumn{5}{c}{Property}\\
   \cline{2-3} \cline{5-9}
   Local Peak & R.A. (J2000) & Decl. (J2000) & & $T_{\rm mb, peak}$ & noise rms & Integrated Intensity & $V_{\rm LSR, peak}$ & $\Delta V_{\rm peak}$\\
   ID Number & (h:m:s) & (d:$^{\prime}$:$^{\prime \prime}$) & & [K] &  [K] & [K km s$^{-1}$] & [km s$^{-1}$] & [km s$^{-1}$]\\
   (1)&(2)&(3)&&(4)&(5)&(6)&(7)&(8)\\ }
\startdata
   \multicolumn{9}{c}{N48}\\
   \hline
\phantom{0}1	&	5:25:47.9	&	$-$66:13:58.9	&&	5.5\phantom{0}	&	0.20 	&	30\phantom{.0}	&	286.0 	&	\phantom{0}5.2 	\\
\phantom{0}2	&	5:25:09.8	&	$-$66:14:48.9	&&	3.6\phantom{0}	&	0.32 	&	28\phantom{.0}	&	291.9 	&	\phantom{0}7.6 	\\
\phantom{0}3	&	5:26:25.8	&	$-$66:10:18.9	&&	2.8\phantom{0}	&	0.31 	&	23\phantom{.0}	&	298.2 	&	\phantom{0}7.9 	\\
\phantom{0}4	&	5:26:02.7	&	$-$66:12:28.9	&&	5.9\phantom{0}	&	0.24 	&	23\phantom{.0}	&	287.6 	&	\phantom{0}3.5 	\\
\phantom{0}5	&	5:25:41.3	&	$-$66:15:08.9	&&	4.3\phantom{0}	&	0.25 	&	19\phantom{.0}	&	291.9 	&	\phantom{0}3.9 	\\
\phantom{0}6	&	5:25:29.7	&	$-$66:16:48.9	&&	4.5\phantom{0}	&	0.38 	&	18\phantom{.0}	&	285.0 	&	\phantom{0}3.8 	\\
\phantom{0}7	&	5:26:06.0	&	$-$66:09:18.9	&&	4.1\phantom{0}	&	0.25 	&	13\phantom{.0}	&	285.5 	&	\phantom{0}3.0 	\\
\phantom{0}8	&	5:25:34.7	&	$-$66:16:08.9	&&	2.8\phantom{0}	&	0.30 	&	11\phantom{.0}	&	292.0 	&	\phantom{0}3.8 	\\
\phantom{0}9	&	5:25:34.7	&	$-$66:17:08.9	&&	4.1\phantom{0}	&	0.36 	&	11\phantom{.0}	&	289.3 	&	\phantom{0}2.5 	\\
10	&	5:25:44.6	&	$-$66:11:48.9	&&	2.5\phantom{0}	&	0.29 	&	11\phantom{.0}	&	279.4 	&	\phantom{0}3.5 	\\
11	&	5:24:58.3	&	$-$66:13:58.9	&&	1.3\phantom{0}	&	0.36 	&	\phantom{0}9.5 		&	291.3 	&	\phantom{0}5.3 	\\
12	&	5:25:57.8	&	$-$66:14:58.9	&&	2.8\phantom{0}	&	0.30 	&	\phantom{0}9.0 		&	288.8 	&	\phantom{0}2.8 	\\
13	&	5:25:39.7	&	$-$66:18:38.9	&&	0.9\phantom{0}	&	0.33 	&	\phantom{0}8.7 		&	287.5 	&	11\phantom{.0}\\
14	&	5:25:24.7	&	$-$66:13:18.9	&&	1.8\phantom{0}	&	0.35 	&	\phantom{0}7.4 		&	297.9 	&	\phantom{0}3.9 	\\
15	&	5:25:21.4	&	$-$66:13:48.9	&&	2.7\phantom{0}	&	0.32 	&	\phantom{0}7.1 		&	297.8 	&	\phantom{0}2.3 	\\
16	&	5:25:46.2	&	$-$66:10:38.9	&&	1.2\phantom{0}	&	0.32 	&	\phantom{0}6.8 		&	294.0 	&	\phantom{0}4.7 	\\
17	&	5:25:29.7	&	$-$66:12:58.9	&&	2.0\phantom{0}	&	0.23 	&	\phantom{0}6.5 		&	298.5 	&	\phantom{0}3.3 	\\
18	&	5:25:59.4	&	$-$66:10:48.9	&&	1.2\phantom{0}	&	0.32 	&	\phantom{0}5.0 		&	295.4 	&	\phantom{0}3.7 	\\
19	&	5:25:33.0	&	$-$66:08:48.9	&&	1.3\phantom{0}	&	0.32 	&	\phantom{0}4.9 		&	298.2 	&	\phantom{0}2.8 	\\
20	&	5:25:18.1	&	$-$66:12:18.9	&&	0.8\phantom{0}	&	0.29 	&	\phantom{0}4.1 		&	283.0 	&	\phantom{0}4.7 	\\
   \hline
   \multicolumn{9}{c}{N49}\\
   \hline
\phantom{0}1	&	5:26:18.2 	&	$-$66:02:53.0	&&	4.5\phantom{0}	&	0.31 	&	23\phantom{.0}	&	285.4	&	\phantom{0}4.9 	\\
\phantom{0}2	&	5:26:16.6	&	$-$66:01:33.0	&&	2.9\phantom{0}	&	0.42 	&	15\phantom{.0}	&	286.8	&	\phantom{0}4.4 	\\
\phantom{0}3	&	5:26:19.9	&	$-$66:02:53.0	&&	2.5\phantom{0}	&	0.30 	&	\phantom{0}7.2		&	292.3	&	\phantom{0}2.7 	\\
\enddata
\tablecomments{Col.(1):ID numbers of local peaks. Col.(2)-(3): Positions of observed point of local peaks.  Col.(4)-(8): Observed properties of the $^{12}$CO($J$=3--2) spectra of the local peaks derived by a single Gaussian fitting for the spectrum. Peak main beam temperature T$_{\rm mb, peak}$, rms of noise per 0.44 km s$^{-1}$ channel of the spectra, integrated intensity, V$_{\rm LSR}$ at the peak of the spectra, and FWHM line width are shown. \label{tab_peak}}
\end{deluxetable}


\begin{deluxetable}{llcccccccc}[ht]
\tablecolumns{10}
\tablewidth{0pc}
\tabletypesize{\scriptsize}
\tablecaption{Parameters of $^{12}$CO($J=$3--2) clumps}
\tablehead{
   &&\multicolumn{2}{c}{Peak Position}& &\multicolumn{5}{c}{Physical Properties}\\
   \cline{3-4} \cline{6-10}
   Clump & Local Peak &  R.A. (J2000) & Decl. (J2000) & & $R_{\rm deconv}$ & $V_{\rm LSR,clump}$ & $\Delta V _{\rm clump}$ &
   $M_{\rm vir}$ & $L_{{\rm CO}(J=3-2)}$\\
   ID Number & ID Number & (h:m:s) & (d:$^{\prime}$:$^{\prime \prime}$) & & [pc] & [km s$^{-1}$] & [km s$^{-1}$] &
   [$\times 10^{4} {\rm M}_{\odot}$] & [$\times 10^{3}{\rm K\;km\;s}^{-1}{\rm pc}^2$]\\
   (1)&(2)&(3)&(4)&&(5)&(6)&(7)&(8)&(9) }
\startdata
   \multicolumn{10}{c}{N48}\\
   \hline
1	&	1	&	5:25:47.9	&	$-$66:13:58.9	&&	3.5	&	286.6	&	5.1	&	1.7\phantom{0}	&	1.7\phantom{0}	\\
2	&	2	&	5:25:09.8	&	$-$66:14:48.9	&&	6.6	&	292.4	&	8.6	&	9.3\phantom{0}	&	3.5\phantom{0}	\\
3	&	3	&	5:26:25.8	&	$-$66:10:18.9	&&	3.8	&	298.1	&	7.6	&	4.2\phantom{0}	&	1.3\phantom{0}	\\
4	&	4	&	5:26:02.7	&	$-$66:12:28.9	&&	4.4	&	287.5	&	3.9	&	1.3\phantom{0}	&	1.7\phantom{0}	\\
5	&	5	&	5:25:41.3	&	$-$66:15:08.9	&&	3.9	&	291.9	&	3.5	&	0.91		&	1.2\phantom{0}	\\
6	&	6	&	5:25:29.7	&	$-$66:16:48.9	&&	2.9	&	285.1	&	4.0 	&	0.88		&	1.0\phantom{0}	\\
7	&	7	&	5:26:06.0	&	$-$66:09:18.9	&&	7.0	&	285.1	&	3.3	&	1.4\phantom{0}	&	1.9\phantom{0}	\\
8	&	8	&	5:25:34.7	&	$-$66:16:08.9	&&	5.4	&	291.7	&	3.7	&	1.4\phantom{0}	&	1.4\phantom{0}	\\
9	&	9	&	5:25:34.7	&	$-$66:17:08.9	&&	2.9	&	289.3	&	3.2	&	0.56		&	0.53		\\
10	&	10	&	5:25:44.6	&	$-$66:11:48.9	&&	5.8	&	280.6	&	5.6	&	3.5\phantom{0}	&	1.0\phantom{0}	\\
11	&	11	&	5:24:58.3	&	$-$66:13:58.9	&&	2.8	&	291.6	&	6.1	&	2.0\phantom{0}	&	0.42		\\
12	&	12	&	5:25:57.8	&	$-$66:14:58.9	&&	1.2	&	288.7	&	2.7	&	0.17		&	0.28		\\
13	&	14	&	5:25:24.7	&	$-$66:13:18.9	&&	1.7	&	297.9	&	3.1	&	0.31		&	0.27		\\
14	&	15	&	5:25:21.4	&	$-$66:13:48.9	&&	3.5	&	297.5	&	3.1	&	0.64		&	0.44		\\
15	&	16	&	5:25:46.2	&	$-$66:10:38.9	&&	2.3	&	294.0 	&	5.0	&	1.1\phantom{0}	&	0.29		\\
16	&	17	&	5:25:29.7	&	$-$66:12:58.9	&&	5.4	&	298.0 	&	3.7	&	1.4\phantom{0}	&	0.64		\\
17	&	18	&	5:25:59.4	&	$-$66:10:48.9	&&	3.2	&	295.3	&	4.5	&	1.2\phantom{0}	&	0.40 		\\
18	&	19	&	5:25:33.0	&	$-$66:08:48.9	&&	3.5	&	298.1	&	2.8	&	0.52		&	0.25		\\
\hline
\multicolumn{10}{c}{N49}\\
\hline
1	&	1	&	5:26:18.2 	&	$-$66:02:53.0	&&	6.6	&	285.2	&	4.2	&	2.2\phantom{0}	&	3.0\phantom{0}	\\
2	&	2	&	5:26:16.6	&	$-$66:01:33.0	&&	8.4	&	287.3	&	4.1	&	2.7\phantom{0}	&	2.7\phantom{0}	\\
3	&	3	&	5:26:19.9	&	$-$66:02:53.0	&&	3.9	&	292.2	&	2.6	&	0.5\phantom{0}		&	0.58		\\
\enddata
\tablecomments{Col.(1): ID numbers of $^{12}$CO($J$=3--2) clumps. Col.(2):ID numbers of $^{12}$CO($J$=3--2) local peaks corresponding to the clump. Col.(3)--(4): Positions of observed point of local peak. Col.(5)--(9): Observed physical properties of $^{12}$CO($J$=3--2) clumps. The deconvolution radius, $R_{\rm deconv}$, the velocity at the spectrum peak, $V_{\rm LSR,clump}$, the FWHM line width for the composite spectra within the clumps, $\Delta V _{\rm clump}$, the virial mass, $M_{\rm vir}$, and the $^{12}$CO($J$=3--2) luminosity of the clump, $L_{{\rm CO}(J=3-2)}$ are shown. \label{tab_clump}}
\end{deluxetable}


\begin{deluxetable}{lccccccccccc}
\tablecolumns{12}
\tablewidth{0pc}
\tabletypesize{\scriptsize}
\tablecaption{Clump properties derived by $^{13}$CO($J$=1--0) transition}
\tablehead{
	&\multicolumn{3}{c}{Clump averaged $^{13}$CO($J$=1-0) properties} &&\multicolumn{5}{c}{Physical Properties of $^{13}$CO($J$=1-0) clumps}\\
   \cline{2-4} \cline{6-10}
   Clump & $T_{\rm mb,clump}$ & $V_{\rm LSR ,clump}$ 
    & $\Delta V_{\rm clump}$  && $R_{\rm deconv}$ & $M_{\rm vir}$ 
    & $\tau$  & $N({\rm H}_{2})$ & $M_{\rm clump}$ && $M_{\rm vir}$/$M_{\rm clump}$\\
   ID Number & [K] & [km s$^{-1}$] & [km s$^{-1}$] && [pc] & [$\times 10^{4} M_{\odot}$] &
   & [$\times 10^{21}$ cm$^{-2}$] & [$\times 10^{4} {\rm M}_{\odot}$]&&\\
   (1)&(2)&(3)&(4)&&(5)&(6)&(7)&(8)&(9)&&(10)\\}
\startdata
   \multicolumn{12}{c}{N48}\\
   \hline
1	&	0.26	&	286.4	&	3.8	&&	3.5	&\phantom{0}0.96		&	0.030\phantom{0}	&	3.8\phantom{0}	&	0.63		&&	1.5\phantom{0}	\\
2	&	0.14	&	292.8	&	9.4	&&	6.6	&		11\phantom{.00}	&	0.016\phantom{0}	&	3.4\phantom{0}	&	2.0\phantom{0}	&&	5.5\phantom{0}	\\
3	&	0.06	&	298.2	&	4.7	&&	3.8	&\phantom{0}1.6\phantom{0}	&	0.0068		&	1.1\phantom{0}	&	0.21		&&	7.6\phantom{0}	\\
4	&	0.24	&	287.9	&	2.5	&&	4.4	&\phantom{0}0.52		&	0.028\phantom{0}	&	2.1\phantom{0}	&	0.55		&&	0.95		\\
5	&	0.16	&	292.1	&	2.7	&&	3.9	&\phantom{0}0.54		&	0.018\phantom{0}	&	1.4\phantom{0}	&	0.29		&&	1.9\phantom{0}	\\
6	&	0.11	&	285.0 	&	4.2	&&	2.9	&\phantom{0}0.97		&	0.013\phantom{0}	&	2.1\phantom{0}	&	0.24		&&	4.0\phantom{0}	\\
7	&	0.16	&	285.4	&	3.5	&&	7.0 	&\phantom{0}1.6\phantom{0}	&	0.018\phantom{0}	&	1.6\phantom{0}	&	1.1\phantom{0}	&&	1.5\phantom{0}	\\
8	&	0.14	&	291.8	&	2.2	&&	5.4	&\phantom{0}0.50 		&	0.016\phantom{0}	&	0.91		&	0.36		&&	1.4\phantom{0}	\\
10	&	0.10 	&	280.9	&	3.9	&&	5.8	&\phantom{0}1.7\phantom{0}	&	0.011\phantom{0}	&	1.2\phantom{0}	&	0.54		&&	3.1\phantom{0}	\\
14	&	0.16	&	297.5	&	1.9	&&	3.5	&\phantom{0}0.24		&	0.018\phantom{0}	&	0.97		&	0.16		&&	1.5\phantom{0}	\\
\hline																									
\multicolumn{12}{c}{N49}\\																							
\hline																									
1	&	0.47	&	285.0 	&	3.7	&&	6.6	&\phantom{0}1.7\phantom{0}	&	0.055\phantom{0}	&	5.3\phantom{0}	&	3.1\phantom{0}	&&	0.55		\\
2	&	0.34	&	287.0 	&	3.9	&&	8.4	&\phantom{0}2.4\phantom{0}	&	0.039\phantom{0}	&	4.1\phantom{0}	&	3.9\phantom{0}	&&	0.62		\\
3	&	0.23	&	291.9	&	2.0 	&&	3.9	&\phantom{0}0.30 		&	0.026\phantom{0}	&	1.5\phantom{0}	&	0.31		&&	0.97		\\
\enddata
\tablecomments{Col.(1):The $^{12}$CO($J$=3--2) clump ID numbers which have been used to determine the boundary of the $^{13}$CO($J$=1--0) clump. Col.(2)--(4): Properties of the clump averaged $^{13}$CO($J$=1--0) spectra derived by a single Gaussian fitting for the spectrum. The peak main beam temperature, $T_{\rm mb,clump}$, the velocity at the spectrum peak, $V_{\rm LSR ,clump}$, and the FWHM line width, $\Delta V_{\rm clump}$, are shown. Col.(5)--(9): Physical properties of the $^{13}$CO($J$=1--0) clumps, the deconvolution radius, $R_{\rm deconv}$, the virial mass, $M_{\rm vir}$, the optical depth, $\tau$, the column density of molecular hydrogen, $N({\rm H}_{2})$, and the LTE mass of $^{13}$CO($J$=1--0) clump, $M_{\rm clump}$, are shown. Col.(10): The ratios of $^{13}$CO($J$=1--0) clump virial mass and LTE mass.\label{tab_13CO}}
\end{deluxetable}


\begin{deluxetable}{llcccrrcc}
\tablecolumns{9}
\tablewidth{0pc}
\tabletypesize{\scriptsize}
\tablecaption{Parameters for LVG analysis}
\tablehead{
   & & \multicolumn{3}{c}{$T _{\rm mb}$} && \multicolumn{2}{c}{Line Ratio}&  \\
\cline{3-5} \cline{7-8} 
 Region &  Clump   & $^{12}$CO($J$=3--2) & $^{12}$CO($J$=1--0) & $^{13}$CO($J$=1--0)  && \multicolumn{1}{c}{$R_{3-2/1-0}$} & $R_{12/13}$ & $dv/dr$ \\
  Name & No. & [K]& [K]& [K]&& & & [km s$^{-1}$] \\
  (1) & (2) & (3) & (4) & (5) && \multicolumn{1}{c}{(6)} & (7) & (8) \\}
\startdata
N48.....&1&3.3\phantom{0}$\pm$0.2&3.2$\pm$0.2&0.31$\pm$0.04&&1.0\phantom{0}$\pm$0.1\phantom{0}&10\phantom{.0}$\pm$1\phantom{.0}\phantom{..}&0.7\\
&2&2.5\phantom{0}$\pm$0.2&3.3$\pm$0.2&0.19$\pm$0.03&&0.76$\pm$0.08&17\phantom{.0}$\pm$3\phantom{.0}\phantom{..}&0.7\\
&3&1.7\phantom{0}$\pm$0.2&1.8$\pm$0.1&0.11$\pm$0.03&&0.94$\pm$0.12&16\phantom{.0}$\pm$5\phantom{.0}\phantom{..}&1.0\\
&4&3.6\phantom{0}$\pm$0.3&2.7$\pm$0.2&0.30$\pm$0.04&&1.3\phantom{0}$\pm$0.1\phantom{0}&\phantom{0}9.0$\pm$1.4\phantom{..}&0.4\\
&5&2.8\phantom{0}$\pm$0.2&2.1$\pm$0.2&0.19$\pm$0.03&&1.3\phantom{0}$\pm$0.2\phantom{0}&11\phantom{.0}$\pm$2\phantom{.0}\phantom{..}&0.4\\
&6&2.3\phantom{0}$\pm$0.2&2.0$\pm$0.2&0.14$\pm$0.04&&1.2\phantom{0}$\pm$0.2\phantom{0}&14\phantom{.0}$\pm$4\phantom{.0}\phantom{..}&0.7\\
&7&2.7\phantom{0}$\pm$0.2&2.8$\pm$0.2&0.21$\pm$0.04&&0.96$\pm$0.10&13\phantom{.0}$\pm$3\phantom{.0}\phantom{..}&0.2\\
&8&1.7\phantom{0}$\pm$0.2&1.7$\pm$0.1&0.17$\pm$0.03&&1.0\phantom{0}$\pm$0.1\phantom{0}&10\phantom{.0}$\pm$2\phantom{.0}\phantom{..}&0.3\\
&14&1.3\phantom{0}$\pm$0.2&2.2$\pm$0.2&0.20$\pm$0.04&&0.59$\pm$0.11&11\phantom{.0}$\pm$2\phantom{.0}\phantom{..}&0.4\\
N49.....&1&3.6\phantom{0}$\pm$0.3&4.8$\pm$0.3&0.56$\pm$0.06&&0.75$\pm$0.08&\phantom{0}8.6$\pm$1.1\phantom{..}&0.3\\
&2&2.1\phantom{0}$\pm$0.2&3.2$\pm$0.2&0.33$\pm$0.06&&0.66$\pm$0.07&\phantom{0}9.7$\pm$1.9\phantom{..}&0.2\\
&3&1.4\phantom{0}$\pm$0.2&1.6$\pm$0.2&0.30$\pm$0.05&&0.88$\pm$0.17&\phantom{0}5.3$\pm$1.1\phantom{..}&0.3\\
\hline
30Dor.....&1&3.7\phantom{0}$\pm$0.5&1.8$\pm$0.4&0.13$\pm$0.03&&2.1\phantom{0}$\pm$0.5\phantom{0}&14\phantom{.0}$\pm$4\phantom{.0}\phantom{..}&0.9\\
&3&2.0\phantom{0}$\pm$0.3&1.5$\pm$0.3&0.13$\pm$0.03&&1.3\phantom{0}$\pm$0.3\phantom{0}&12\phantom{.0}$\pm$4\phantom{.0}\phantom{..}&0.9\\
&4&2.3\phantom{0}$\pm$0.4&1.7$\pm$0.3&0.13$\pm$0.03&&1.4\phantom{0}$\pm$0.3\phantom{0}&13\phantom{.0}$\pm$4\phantom{.0}\phantom{..}&0.5\\
N159.....&1&8.6\phantom{0}$\pm$1.2&6.2$\pm$1.2&0.80$\pm$0.16&&1.4\phantom{0}$\pm$0.3\phantom{0}&\phantom{0}7.8$\pm$2.2\phantom{..}&0.9\\
&2&6.1\phantom{0}$\pm$0.8&4.0$\pm$0.8&0.44$\pm$0.09&&1.5\phantom{0}$\pm$0.4\phantom{0}&\phantom{0}9.1$\pm$2.6\phantom{..}&0.5\\
&4&3.8\phantom{0}$\pm$0.5&4.5$\pm$0.9&0.72$\pm$0.14&&0.84$\pm$0.20&\phantom{0}6.3$\pm$1.7\phantom{..}&0.4\\
N166.....&1&3.3\phantom{0}$\pm$0.7&4.0$\pm$0.8&--\footnotemark[1]&&0.83$\pm$0.17&11\phantom{.0}$\pm$2\phantom{.0}\footnotemark[1]&0.3\\
&3&1.6\phantom{0}$\pm$0.3&2.3$\pm$0.5&--\footnotemark[1]&&0.70$\pm$0.15&13\phantom{.0}$\pm$3\phantom{.0}\footnotemark[1]&0.3\\
&4&0.98$\pm$0.2&1.3$\pm$0.3&--\footnotemark[1]&&0.75$\pm$0.17&16\phantom{.0}$\pm$3\phantom{.0}\footnotemark[1]&0.4\\
N206.....&1&2.6\phantom{0}$\pm$0.5&3.1$\pm$0.6&--\footnotemark[1]&&0.84$\pm$0.16&14\phantom{.0}$\pm$3\phantom{.0}\footnotemark[1]&0.5\\
&2&1.3\phantom{0}$\pm$0.3&2.1$\pm$0.4&--\footnotemark[1]&&0.62$\pm$0.12&\phantom{0}9.8$\pm$2.0\footnotemark[1]&0.4\\
N206D.....&1&3.3\phantom{0}$\pm$0.5&4.6$\pm$0.9&0.85$\pm$0.17&&0.72$\pm$0.18&\phantom{0}5.4$\pm$1.5\phantom{..}&0.3\\
GMC225.....&1&1.5\phantom{0}$\pm$0.2&3.4$\pm$0.7&0.55$\pm$0.11&&0.44$\pm$0.11&\phantom{0}6.2$\pm$1.8\phantom{..}&0.2\\
&3&1.2\phantom{0}$\pm$0.2&2.8$\pm$0.6&--\footnotemark[1]&&0.43$\pm$0.09&\phantom{0}6.7$\pm$1.3\footnotemark[1]&0.3\\
\enddata
\tablecomments{Col.(1)--(2): Names of the region and ID numbers of the clumps that we performed LVG analysis. Col.(3)--(5): Peak main beam temperatures of $^{12}$CO($J$=3--2), $^{12}$CO($J$=1--0) , $^{13}$CO($J$=1--0) lines. Col.(6)--(7): Line intensity ratios that we have used in LVG analysis. Ratio of $^{12}$CO($J$=3--2) and $^{12}$CO($J$=1--0), $R_{3-2/1-0}$, and ratio of $^{12}$CO($J$=1--0) and $^{13}$CO($J$=1--0), $R_{12/13}$, are shown. Col.(8): Averaged velocity gradients of the clumps. \label{tab_lvgpara}}
\tablenotetext{a}{The value quoted from Minamidani et al. (2008) as an intensity ratio $R_{12/13}$.}
\end{deluxetable}


\begin{deluxetable}{llccccccccc}
\tablecolumns{11}
\tablewidth{0pc}
\tabletypesize{\scriptsize}
\tablecaption{Results of LVG analysis}
\tablehead{
 Region &  Clump   &&  \multicolumn{2}{c}{$n$(H$_{2}$) [cm$^{-3}$]} && \multicolumn{2}{c}{$T _{\rm kin}$ [K]} & min. $\chi ^{2}$ && $F _{\rm 24 \mu m}$\\
 \cline{4-5} \cline{7-8} 
  Name & No. && $\chi ^{2}<3.84$ & min. $\chi ^{2}$ && $\chi ^{2}<3.84$ & min. $\chi ^{2}$ &&& 10$^{-12}$ [erg s$^{-1}$ cm$^{-2}$]  \\
  (1) & (2) && (3) & (4) && (5) & (6) & (7)&& (8)\\}
\startdata
N48.....&1&&$10^{3.3}$--$10^{5.2}$&$10^{5.0}$&&$>$46&191&\phantom{0}0.023\phantom{00}&&\phantom{00}99\phantom{.0}\\
&2&&$10^{2.8}$--$10^{3.3}$&$10^{3.0}$&&$46-183$&91&\phantom{0}0.053\phantom{00}&&\phantom{000}8.4\\
&3&&$10^{3.0}$--$10^{5.5}$&$10^{3.4}$&&$>$34&\phantom{0}72&\phantom{0}0.035\phantom{00}&&\phantom{00}12\phantom{.0}\\
&4&&---&&&---&---&10\phantom{.00000}&&\phantom{00}31\phantom{.0}\\
&5&&$10^{3.3}$--$10^{5.8}$&$10^{4.0}$&&$>$79&126&\phantom{0}2.0\phantom{0000}&&\phantom{0}110\phantom{.0}\\
&6&&$10^{3.1}$--$10^{5.2}$&$10^{4.1}$&&$>$52&126&\phantom{0}0.33\phantom{000}&&\phantom{00}41\phantom{.0}\\
&7&&$10^{2.7}$--$10^{4.6}$&$10^{3.1}$&&$>$84&191&\phantom{0}0.00030&&\phantom{00}26\phantom{.0}\\
&8&&$10^{3.1}$--$10^{4.9}$&$10^{3.9}$&&$>$61&120&\phantom{0}0.0039\phantom{0}&&\phantom{00}43\phantom{.0}\\
&14&&$10^{2.6}$--$10^{3.1}$&$10^{2.9}$&&$23-96$\phantom{0}&\phantom{0}50&\phantom{0}0.070\phantom{00}&&\phantom{00}11\phantom{.0}\\
N49.....&1&&$10^{2.7}$--$10^{4.8}$&$10^{3.0}$&&$>$27\phantom{0}&\phantom{0}55&\phantom{0}0.010\phantom{00}&&\phantom{00}15\phantom{.0}\\
&2&&$10^{2.4}$--$10^{4.5}$&$10^{2.7}$&&$>$19&\phantom{0}72&\phantom{0}0.0036\phantom{0}&&\phantom{000}6.1\\
&3&&$10^{2.8}$--$10^{5.3}$&$10^{3.5}$&&$>$13&\phantom{0}50&\phantom{0}0.0054\phantom{0}&&\phantom{00}16\phantom{.0}\\
\hline
30Dor.....&1&&---&---&&---&---&\phantom{0}4.2\phantom{0000}&&5600\phantom{.0}\\
&3&&$10^{2.9}$--$10^{4.7}$&$10^{4.5}$&&$>$51&126&\phantom{0}0.95\phantom{000}&&1900\phantom{.0}\\
&4&&$10^{3.1}$--$10^{5.0}$&$10^{4.1}$&&$>$61&151&\phantom{0}1.1\phantom{0000}&&1200\phantom{.0}\\
N159.....&1&&$10^{3.5}$--$10^{5.6}$&$10^{5.3}$&&$>$35&182&\phantom{0}2.0\phantom{0000}&&2000\phantom{.0}\\
&2&&$10^{3.1}$--$10^{5.3}$&$10^{4.9}$&&$>$38&191&\phantom{0}1.7\phantom{0001}&&2500\phantom{.0}\\
&4&&$10^{2.9}$--$10^{5.4}$&$10^{3.4}$&&$>$15&\phantom{0}44&\phantom{0}0.0041\phantom{0}&&\phantom{00}23\phantom{.0}\\
N166.....&1&&$10^{2.6}$--$10^{4.8}$&$10^{3.1}$&&$>$38&\phantom{0}91&\phantom{0}0.011\phantom{00}&&\phantom{00}11\phantom{.0}\\
&3&&$10^{2.4}$--$10^{3.5}$&$10^{2.9}$&&$>$32&100&\phantom{0}0.058\phantom{00}&&\phantom{000}8.0\\
&4&&$10^{2.4}$--$10^{3.6}$&$10^{2.9}$&&$>$46&132&\phantom{0}0.0041\phantom{0}&&\phantom{00}25\phantom{.0}\\
N206.....&1&&$10^{2.7}$--$10^{5.0}$&$10^{3.1}$&&$>$39&\phantom{0}87&\phantom{0}0.0087\phantom{0}&&\phantom{0}140\phantom{.0}\\
&2&&$10^{2.7}$--$10^{3.3}$&$10^{2.9}$&&$20-87$\phantom{0}&\phantom{0}44&\phantom{0}0.080\phantom{00}&&\phantom{00}51\phantom{.0}\\
N206D.....&1&&$10^{2.7}$--$10^{5.3}$&$10^{3.1}$&&$>$12&\phantom{0}32&\phantom{0}0.0042\phantom{0}&&\phantom{0}100\phantom{.0}\\
GMC225.....&1&&$10^{2.5}$--$10^{2.8}$&$10^{2.7}$&&$13-48$\phantom{0}&\phantom{0}19&\phantom{0}0.24\phantom{000}&&\phantom{000}2.8\\
&3&&$10^{2.6}$--$10^{3.0}$&$10^{2.7}$&&$11-34$\phantom{0}&\phantom{0}19&\phantom{0}0.16\phantom{000}&&\phantom{000}1.9\\
\enddata
\tablecomments{Col.(1)--(2): Names of the regions and ID numbers of the clumps that we performed the LVG analysis. Col.(3)--(6): Derived number density, $n$(H$_2$) cm$^{-3}$, and kinetic temperature, $T_{\rm kin}$, of the clumps are shown. The range of value for that $\chi ^{2}$ is less than 3.84, and the values at the point that minimum $\chi ^2$ are indicated. Col.(7): The values of minimum $\chi ^{2}$. Col.(8): Flux densities of Spitzer 24 $\mu$m from each 45$^{\prime \prime}$ area toward the peak position of the clumps.\label{tab_lvgresults}}
\end{deluxetable}


\begin{table}[ht]
 \caption{H$\alpha$ emission nebulae identified around the N48/N49 regions  \label{tab_Hanebula}}
 \label{tab_Hanebula}
 \begin{center}
 \scalebox{0.9}{
  \begin{tabular}{lcccc}
   \hline \hline
   & \multicolumn{2}{c}{Source Position} & &\\
   \cline{2-3}
   Source Name &  R.A. (J2000) & Decl. (J2000) & Source Type & reference\\
   & (h:m:s) & (d:$^{\prime}$:$^{\prime \prime}$) & & \\
   \hline
   N48-A, DEM L189 & 05:25:49.2 & $-$66:15:08.8 & Emission Nebura & (1),(2)\\
   N48-B, DEM L189 & 05:25:41.1 & $-$66:17:38.1 & \ion{H}{2} region& (1),(2)\\
   N48-C, DEM L189 & 05:25:52.0 & $-$66:14:32\phantom{.0} & \ion{H}{2} region& (1),(2)\\
   DEM L191 & 05:26:07.3 & $-$66:09:26.1 &  \ion{H}{2} region& (2)\\
   N49, DEM L190 & 05:26:01.9 & $-$66:04:59.7 & SNR& (1),(2)\\
   DEM L181  & 05:25:17.2 & $-$66:00:28.5 & HII region& (2)\\
   \hline
  \end{tabular}
  } 
 \end{center}
 \tablerefs{
 (1) \cite{Henize_1956}, 
 (2) \cite{Davies_Meaburn_1976}
 }
\end{table}

\clearpage




\clearpage

\appendix
\section{Complete LVG results}
\begin{figure}[ht]
\epsscale{1.0}
\plotone{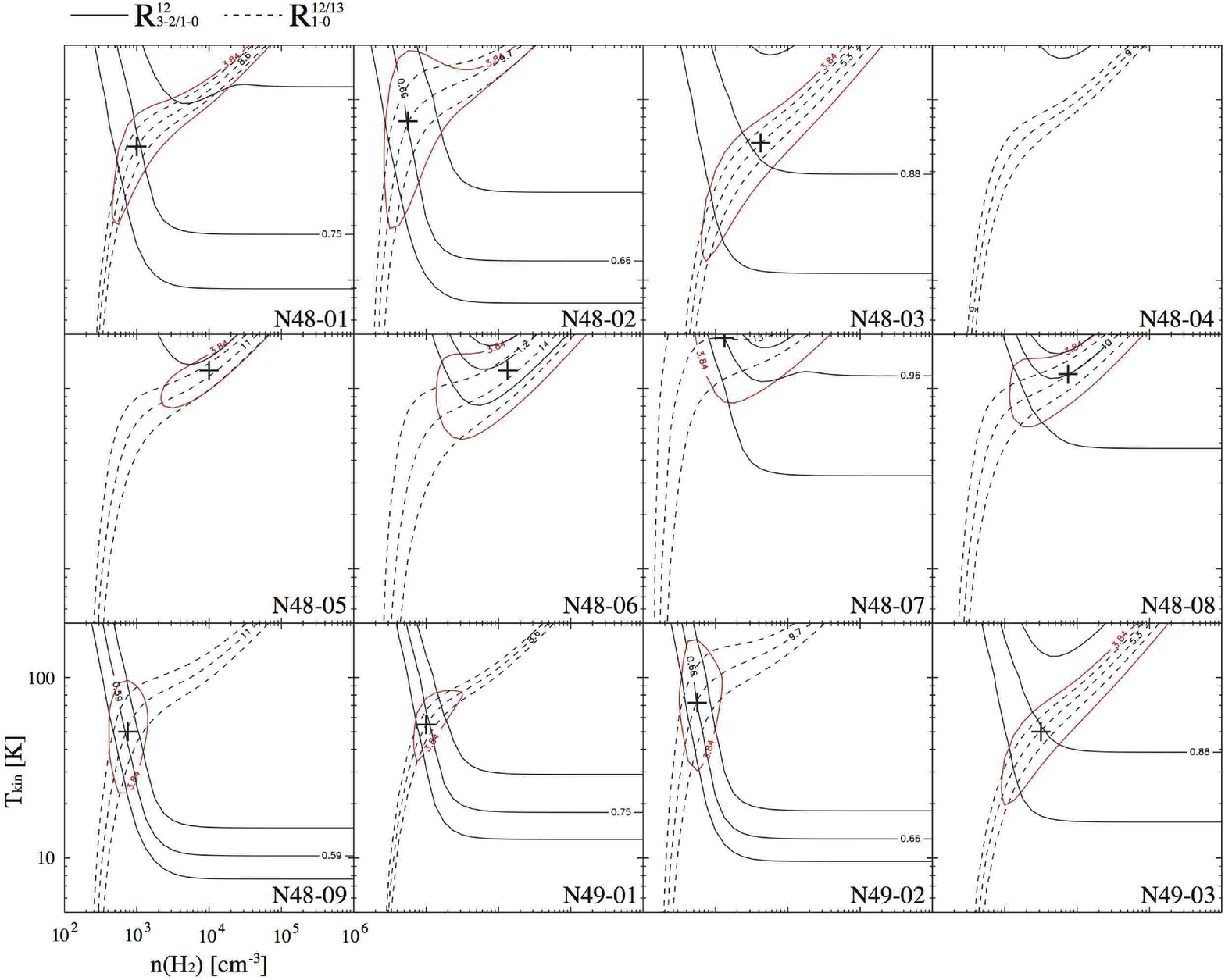}
\caption{The results of the LVG analysis for the N48 and N49 regions. Crosses denote the points of lowest chi-square $\chi ^2$. Red contours indicate $\chi ^2$ = 3.84, which corresponds to the 5\% confidence level of the $\chi ^2$ distribution with one degree of freedom. Black lines show the intensity ratios: $R^{12} _{3-2/1-0}$ (solid lines), $R^{12/13} _{1-0}$ (dashed lines). Each consists of the observed intensity ratios (center) and uncertainty envelopes (outer two lines) that are estimated to be 8\%--31\% for $R^{12} _{3-2/1-0}$ and 10\%--31\% for $R^{12/13} _{1-0}$. \label{fig_nTplot_N4849}}
\end{figure}

\begin{figure}
\epsscale{1.0}
\plotone{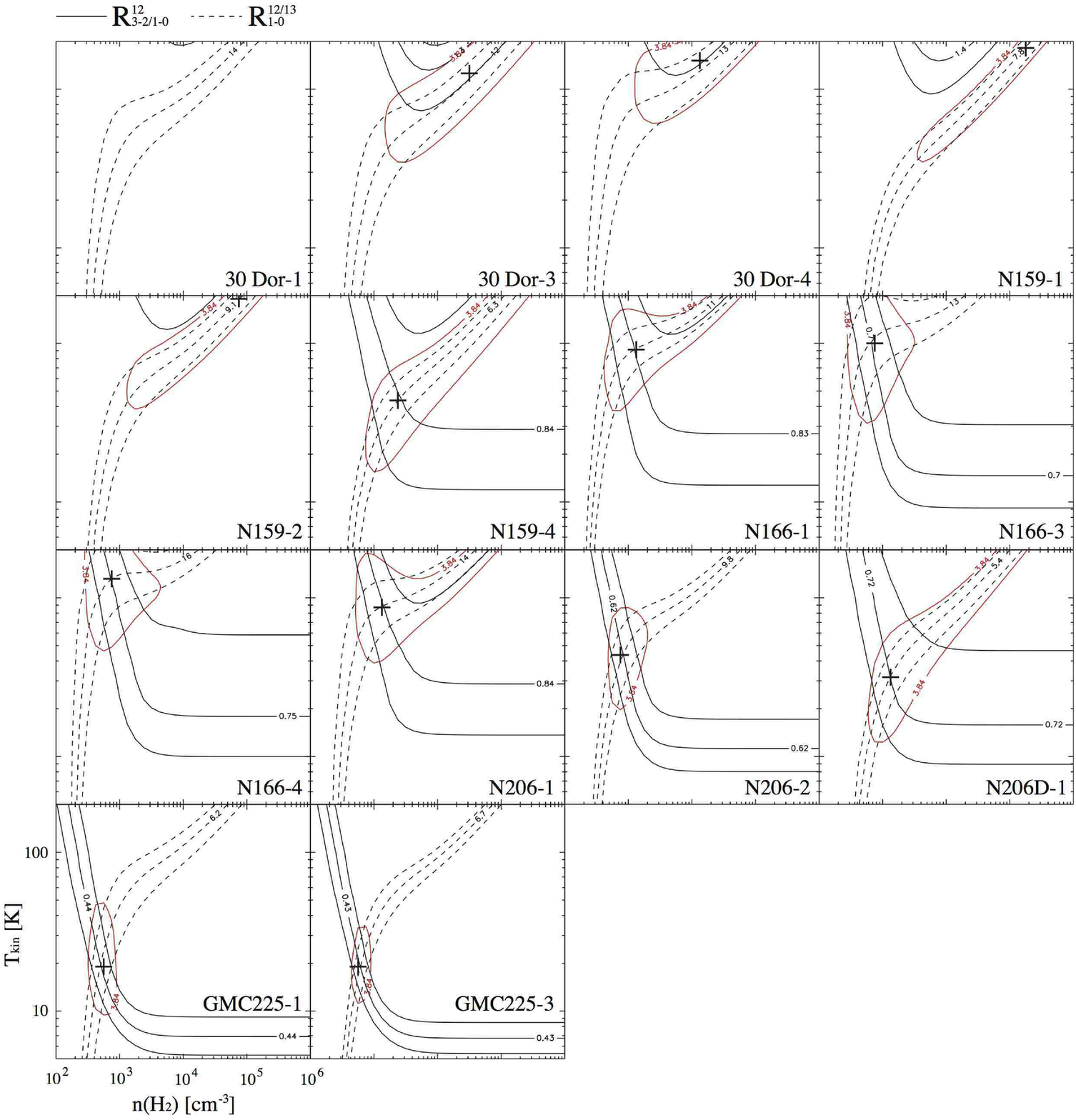}
\caption{The results of the LVG analysis for the M0811 clumps. Details are the same as in Figure \ref{fig_nTplot_N4849}. Uncertainty envelopes (outer two lines) that are estimated to be 20\%--28\% for $R^{12} _{3-2/1-0}$ and 20\%--33\% for $R^{12/13} _{1-0}$. \label{fig_nTplot_M0811}}
\end{figure}

Figures \ref{fig_nTplot_N4849} and \ref{fig_nTplot_M0811} show the resulting contour plots of LVG analysis on the $n({\rm H}_2) - T_{\rm kin}$ plane. Detail of the analysis is described in \S \ref{lvg}. 

\clearpage

\clearpage

\end{document}